\documentclass[11pt]{amsart}
\usepackage{amsbsy,amssymb,amsmath,amsthm,amscd,amsfonts,latexsym,amstext,delarray,
amsmath,graphicx} 
\usepackage[margin=1in]{geometry}
\usepackage{color}
\input xypic

\usepackage[all]{xy}
\usepackage{setspace}

\newtheorem{thm}{Theorem}[section]
\newtheorem{prop}[thm]{Proposition}
\newtheorem{cor}[thm]{Corollary}
\newtheorem{lem}[thm]{Lemma}

\newtheorem{defn}[thm]{Definition}
\newtheorem{rem}[thm]{Remark}

\numberwithin{equation}{section}

\def\F{{\mathbb F}}
\def\Q{{\mathbb Q}}
\def\Z{{\mathbb Z}}
\def\N{{\mathbb N}}
\def\R{{\mathbb R}}
\def\C{{\mathbb C}}

\def\bG{{\mathbb G}}
\def\P{{\mathbb P}}
\def\A{{\mathbb A}}
\def\GL{{\rm GL}}

\def\cA{{\mathcal A}}
\def\cB{{\mathcal B}}
\def\cC{{\mathcal C}}

\def\cF{{\mathcal F}}
\def\cG{{\mathcal G}}
\def\cH{{\mathcal H}}

\def\cL{{\mathcal L}}

\def\cO{{\mathcal O}}

\def\cR{{\mathcal R}}

\def\bG{{\mathbb G}}

\def\bK{{\mathbb K}}

\def\bP{{\mathbb P}}
\def\bQ{{\mathbb Q}}

\def\bT{{\mathbb T}}

\DeclareMathOperator*{\Hom}{Hom}

\title{Endomotives of Toric Varieties}
\author{Zhaorong Jin and Matilde Marcolli}
\address{Mathematics Department, Caltech, 1200 E. California Blvd. Pasadena, CA 91125, USA}
\email{zjin@caltech.edu}
\email{matilde@caltech.edu}
\date{}

\begin{document}

\begin{abstract}
We construct endomotives associated to toric varieties, in terms of the
decomposition of a toric variety into torus orbits and the action of a
semigroup of toric morphisms. We show that the endomotives can be
endowed with time evolutions and we discuss the resulting quantum
statistical mechanical systems. We show that in particular, one can
construct a time evolution related to the logarithmic height function.
We discuss relations to $\F_1$-geometry.
\end{abstract}

\maketitle
\tableofcontents

\section{Introduction}

The notion of endomotive was introduced in \cite{CCM} as
a way to describe, in terms of arithmetic data, the construction
of quantum statistical mechanical systems associated to number theory, 
starting with the prototype Bost--Connes system \cite{BC}. 
Endomotives are algebras obtained from projective limits of Artin 
motives (zero dimensional algebraic varieties) endowed with 
semigroup actions. Out of these algebraic data one obtains 
noncommutative spaces, in the form of semigroup crossed 
product $C^*$-algebras.

\smallskip

There are natural time evolutions on the $C^{*}$-algebra of an endomotive,
and the associated Hamiltonian, partition function and KMS
equilibrium states relate to properties of $L$-functions in cases
arising from number theory, see \cite{CCM}, \cite{CorMa}, \cite{Yal}. 
In particular, endomotives were recently used by Bora Yalkinoglu
to construct arithmetic subalgebras for all the quantum statistical
mechanics associated to number fields, \cite{Yal}, in the context
of the noncommutative geometry approach to the explicit class field
theory problem. It is also known, see \cite{Bor}, \cite{CCM2}, \cite{Man}, \cite{Mar}, \cite{Yal},
that the notion of endomotive is closely related to the notion of
$\Lambda$-rings studied by Borger, in his approach to geometry over
the ``field with one element" $\mathbb{F}_{1}$. 

\smallskip

An interesting question in the theory of endomotives is whether
the construction extends to more general algebro-geometric objects,
besides the cases underlying the construction of quantum statistical
mechanical systems of number fields, that was the main focus in 
\cite{CCM}, \cite{CorMa}, \cite{Yal}. Toric varieties are a natural choice
of a class of varieties for which this question can be addressed. In fact,
toric varieties constitute an important class of algebraic varieties, 
which is sufficiently concrete and well understood (\cite{Cox}, \cite{Ful})
to provide a good testing ground for various constructions. Moreover,
toric varieties play an important role in the theory of $\mathbb{F}_{1}$-geometry.
A first step in the direction of the construction of associated endomotives
already existed, in the form of the multivariable Bost--Connes systems
introduced in \cite{Mar}, which we will interpret here as the simplest
case of our construction for toric varieties, corresponding to the case
where the variety is just a torus $\mathbb{T}^n$.

\smallskip

In this paper, we construct
endomotives for abstract toric varieties, generalizing the existing
constructions of the Bost-Connes system and its generalizations. 

In the rest of this section, we recall the basic notion of an
endomotive, as defined in \cite{CCM} and the main properties
of the associated quantum statistical mechanical systems.
We also recall the torus-cone correspondence for toric
varieties, which will be the basis of our construction. 
We then identify a semigroup of toric morphisms which
we will use in the endomotive data. 

In Section \ref{EndoSec} we describe the construction of
the endomotives of abstract toric varieties. We give two
variants of the construction, which we refer to as the
additive and multiplicative case, which correspond,
respectively, to the abelian part of the endomotive algebra
being a direct sum or a tensor product of contributions from
the single orbits. The direct sum choice is more closely
tied up to the geometry of the variety, as it corresponds
to the decomposition into a union of orbits, with the abelian
part of the endomotive given by the algebra of functions on
a set of algebraic points on the toric variety obtained as
a projective limit over the action of the semigroup.
The tensor product case corresponds instead to regarding the
torus orbits as defining  independent quantum mechanical systems,
so that the resulting partition function will decompose as a
product over orbits. In both cases we obtain Hilbert space
representations of the abelian algebras and of the semigroup,
in such a way that they determine a representation of the
semigroup crossed product algebra. We describe the
explicit generators and relations of the crossed product.
We then describe a general procedure to construct time
evolutions on the algebra with the corresponding Hamiltonians
that are the infinitesimal generators in the given representation.
We describe the group of symmetries and the partition function.
In some especially nice cases, from the point of view of
symmetries of the fan defining the toric variety, we give a more
concrete description of the Hamiltonian and the partition function.

In Section \ref{F1sec}, we relate the endomotives of
toric varieties to $\F_1$-geometry, both in the sense of Borger, \cite{Bor},
via the notion of $\Lambda$-ring, and in the sense of Soul\'e, \cite{Soule}.
We also show that a weaker form of the endomotive construction
can be extended from toric varieties to torified spaces in the sense of
the approach to $\F_1$-geometry of Lorscheid and L\'opez-Pe\~{n}a, \cite{LLP1}.

In Section \ref{HeightSec}, we focus on the case of projective toric varieties,
and in particular on the concrete example of projective spaces. We 
replace a set of distinguished points in the torus orbits used in the
abstract construction of Section \ref{EndoSec} with 
the set of $\mathbb{\bar{Q}}$-algebraic points of the variety with
bounded height and degree over $\mathbb{Q}$, and we describe
a time evolution and covariant representations of the resulting $C^*$-dynamical
system related to the logarithmic height function. We show a variant of the
construction for the case of affine spaces. 

Finally, in Section \ref{GibbsSec}, we discuss briefly the Gibbs equilibrium
states fort the quantum statistical mechanical systems of endomotives of
abstract toric varieties.

\medskip
\subsection{The notion of endomotive}

We recall here briefly the notion of endomotive from \cite{CCM} and
the main properties we will be discussing in the rest of the paper.

\smallskip

The data of an endomotive consist of a projective system $X_{\alpha}$ 
of zero dimensional algebraic varieties over a field $\bK$, 
where $X_{\alpha}={\rm Spec}(A_{\alpha})$,
together with an action by endomorphisms of a semigroup $S$
on the limit $X=\varprojlim_\alpha X_{\alpha}$. In \cite{CCM} the
field $\bK$ is assumed to be a number field.

\smallskip

In \cite{CCM} the semigroup $S$ is assumed to be countably generated
and {\em abelian}, while more general situations with $S$ not necessarily
abelian were discussed, for instance, in \cite{Mar} and will also be
considered here.

\smallskip

At the algebraic level, one associates to the data $(X,S)$ of an
endomotive the algebraic semigroup crossed product algebra 
$A\rtimes S$, while at the analytic level, one considers
the $C^{*}$-algebra $C(X(\bar \bK))\rtimes S$. 

\smallskip

At the algebraic level, one considers the algebraic semigroup crossed
product $\bK$-algebra $A\rtimes S$, where 
$A=\varinjlim_\alpha A_\alpha$ and $X={\rm Spec}(A)$. This is
generated algebraically by elements $a\in A$ and additional
generators $\mu_s$ and $\mu_s^*$, for $s\in S$, satisfying the
relations $\mu_s^* \mu_s=1$, $\mu_s \mu_s^*=\phi_s(1)$, where $\phi_s$
is the endomorphism of $A$ corresponding to $s\in S$, and 
$\mu_{s_1 s_2}=\mu_{s_1}\mu_{s_2}$, $\mu^*_{s_2s_1}=\mu^*_{s_1}\mu^*_{s_2}$,
$\mu_s \, a = \phi_s(a)\, \mu_s$, and $a\, \mu_s^* = \mu_s^* \phi_s(a)$,
for all $s, s_1,s_2 \in S$ and for all $a\in A$.

\smallskip
\subsubsection{Quantum statistical mechanical systems of endomotives}

At the analytic level, one considers the $C^{*}$-algebra $\cA=C(X)\rtimes S$. 
One regards this as the algebra of observables of a quantum statistical
mechanical system, with a time evolution given by a one parameter
family of automorphisms $\sigma: {\mathbb R}\to {\rm Aut}(\cA)$. 

A covariant representation of the $C^*$-dynamical system $(\cA, \sigma_t)$
is a pair $(\pi, H)$ of a representation $\pi: \cA \to \cB(\cH)$ of the algebra by bounded operators
on a Hilbert space $\cH$, and an operator $H$ on $\cH$ (usually unbounded) 
with the property that
\begin{equation}\label{covrep}
\pi(\sigma_t (a)) = e^{-it\, H}\, \pi(a) \, e^{it \, H}.
\end{equation}
One says that $H$ is the Hamiltonian generating the the time evolution $\sigma_t$
in the representation $\pi$.

The typical form of the time evolution considered in \cite{CCM} arises 
from the modular automorphism group $\sigma_t^\varphi$ associated
to a state $\varphi$ determined by a measure on $X$. Here we will give
a construction of time evolutions on endomotives, based more generally
on semigroup homomorphisms $g: S \to \R^*_+$ for which there exists
an associated function $h$, with appropriate scaling properties, so
that the pair $(g,h)$ determine a time evolution and the corresponding
Hamiltonian in an assigned representation (see Proposition \ref{Lemma2.6} below).

\smallskip
\subsubsection{Partition function}

Given a $C^*$-dynamical system $(\cA,\sigma_t)$ and a covariant
representation $(\pi,H)$, the partition function of the system is given by 
\begin{equation}\label{ZpiH}
Z(\beta) = {\rm Tr}(e^{-\beta\, H}).
\end{equation}
In typical situations, there is a sufficiently large real $\beta_0$ such that for
all real $\beta$ with $\beta > \beta_0$ the operator $e^{-\beta H}$ is trace class. 
In physical terms one thinks of the variable $\beta$ as an inverse temperature
(up to the Boltzmann constant).

\smallskip
\subsubsection{Equilibrium states}

A state is a continuous linear functional $\varphi: \cA \to \C$, which
satisfies positivity, $\varphi(a^* a)\geq 0$ for all $a\in \cA$, and
normalized by $\varphi(1)=1$. States are the analog of probability
measures on noncommutative spaces. An equilibrium state of a
$C^*$-dynamical system $(\cA, \sigma_t)$ is  a state satisfying 
$\varphi(\sigma_t(a))=\varphi(a)$ for all $a\in \cA$ and all $t\in \R$.

An important class of equilibrium states on a system $(\cA, \sigma_t)$
are the Gibbs states, which are of the form
\begin{equation}\label{Gibbs2}
\varphi(a) =\frac{{\rm Tr}(\pi(a) \, e^{-\beta H})}{{\rm Tr}(e^{-\beta H})},
\end{equation}
for a covariant representation $(\pi, H)$. These are well defined when
$e^{-\beta H}$ is trace class. A more general class of equilibrium states
generalizing Gibbs states is given by the 
KMS$_\beta$ states (Kubo--Martin--Schwinger). 

\medskip
\subsection{Toric varieties and orbit-cone decomposition}

We recall some basic results from the theory of toric varieties, 
in particular the {\em orbit-cone correspondence}, which we will need to
use extensively in the following. We refer the reader to 
the first three chapters of \cite{Cox} for more details 
and complete proofs. 

\smallskip

Let $N$ be a lattice of rank $d$, and let $M$ be its dual lattice.
The lattices $N$ and $M$ are, respectively, the lattice of 
one-parameter subgroups and the character lattice of the toric variety. 
Let $N_{\mathbb{R}}$ and $M_{\mathbb{R}}$
be the real vector spaces obtained by tensoring $N$ and $M$ with
$\mathbb{R}$, respectively. 

\smallskip

Let $\Sigma$ be a fan in $N_{\mathbb{R}}$,
and $X_{\Sigma}$ be the abstract toric variety associated to $\Sigma$.

\smallskip

Suppose that $\Sigma=\{\sigma_{1},\sigma_{2},...,\sigma_{m}\}$, as a set
of cones. Then, according to the orbit-cone correspondence, we can decompose
$X_{\Sigma}$ as a disjoint union of torus orbits 
\begin{equation}\label{XSigmaorb}
X_{\Sigma} = \overset{m}{\underset{k=1}{\sqcup}}O(\sigma_{k}),
\end{equation}
with the orbits given by 
\begin{equation}\label{Osigma}
O(\sigma_{k})=\{\gamma:\sigma^{\vee}\cap M\rightarrow\mathbb{C}|\gamma(u)\neq0\Longleftrightarrow u\in\sigma^{\bot}\cap M\}\simeq {\rm Hom}_{\mathbb{Z}}(\sigma^{\bot}\cap M,\mathbb{C}^{*})\simeq T_{\sigma^{\bot}\cap M} \, ,
\end{equation}
where $\sigma^{\bot}$ denotes the orthonormal complement of $\sigma$
in $M_{\mathbb{R}}$ and $\sigma^{\vee}$ the dual cone of $\sigma$.

\smallskip

Each torus orbit $O(\sigma_{k})$ contains a distinguished point
$\gamma_{k}$ given by 
\begin{equation}
\gamma_k \, : \,  u\in\sigma_{k}^{\vee}\cap M\mapsto\begin{cases}
1 & \mathrm{if}\, u\in\sigma_{k}^{\bot}\cap M\\
0 & \mathrm{otherwise.}
\end{cases}\label{eq:1}
\end{equation}

\smallskip

The orbit-cone correspondence  
will be the foundation of our construction of endomotives.

\smallskip
\subsection{Semigroup of an abstract toric variety}\label{Ssec}

We introduce here a semigroup of toric
morphisms, associated to an abstract toric variety, which is
defined directly in terms of the orbit-cone decomposition, and
which will provide the semigroup for the endomotive construction.

\smallskip

We call a $\mathbb{Z}$-linear transformation $\phi$ of
$N$ compatible with the fan $\Sigma$ if, for any cone $\sigma_{j}$
in $\Sigma$, there exists a cone $\sigma_{k}$ in $\Sigma$ such
that $\phi_{\mathbb{R}}(\sigma_{j})\subseteq\sigma_{k}$, 
where $\phi_{\mathbb{R}}=\mbox{\ensuremath{\phi}}\otimes1$
is the induced linear map on $N_{\mathbb{R}}$. For each such map
$\phi$ and each cone $\sigma_{j}$ in $\Sigma$, there exists a unique
cone $\sigma_{k}$ in $\Sigma$ such that 
\begin{equation}\label{Relinteq}
\phi_{\mathbb{R}}({\rm Relint}(\sigma_{j}))\subseteq {\rm Relint}(\sigma_{k}).
\end{equation}
Thus $\phi$ can be regarded as a self-map on $\{\sigma_{1},\sigma_{2},...,\sigma_{m}\}$.

\smallskip

Let $S_{0}$ denote the set of all nonsingular linear transformations of
$N$ that are compatible with the fan $\Sigma$ and bijective as a
self-map on the cones in $\Sigma$. Then $S_{0}$ is a semigroup under
composition (matrix multiplication if we identify the elements 
with $d\times d$-matrices). 

\smallskip

A standard fact about toric varieties implies that
each $\phi$ induces a toric morphism $\bar{\phi}$ of $X_{\Sigma}$
in the following way: for each cone $\sigma_{j}$ in $\Sigma$, let 
$\sigma_{k}$ be the cone in $\Sigma$ such that $\phi$ maps ${\rm Relint}(\sigma_{j})$
into ${\rm Relint}(\sigma_{k})$; then we have a map of 
corresponding torus orbits 
\begin{equation}\label{maporbs}
\bar{\phi}:O(\sigma_{k})\simeq {\rm Hom}_{\mathbb{Z}}(\sigma_{k}^{\bot}\cap M,\mathbb{C}^{*})\rightarrow O(\sigma_{j})\simeq {\rm Hom}_{\mathbb{Z}}(\sigma_{j}^{\bot}\cap M,\mathbb{C}^{*})
\end{equation}
\begin{equation}\label{eq:2}
f_{k}\mapsto f_{k}\circ\phi^{T}, 
\end{equation}
where $\phi^{T}$ is the transposed matrix of $\phi$. To simplify
notation, we denote by $S$ the semigroup consisting of the transposes of
the matrices in $S_{0}$. Then $S$ acts on $X_{\Sigma}$ by precomposition. 

\smallskip

\section{Endomotives of abstract toric varieties}\label{EndoSec}

In this section we present our main construction of endomotives for abstract
toric varieties. The algebra of observables and the time evolution are obtained
from the orbit-cone decomposition and the semigroup introduced above, and
from a class of functions with suitable scaling properties under composition
with the toric morphisms of the semigroup.

\medskip
\subsection{Multivariable Bost--Connes systems}

We first recall briefly the case of a single torus, which corresponds
to the ``multivariable Bost--Connes systems" introduced in \cite{Mar}.
One considers in that case a variety that is an $n$-dimensional 
algebraic torus $\bT^n=(\bG_m)^n$. The semigroup $S_n$ is given
by $n\times n$-matrices with integer entries and positive determinant, 
$S_n=M_n(\Z)^+$, with the action on $\bT^n$ given by
\begin{equation}\label{Cremonaaction}
t=(t_j)_{j=1}^n \stackrel{\alpha}{\mapsto} t^\alpha=(t^\alpha_i), \ \text{ where } \
t_i^\alpha =\prod_j t_j^{\alpha_{ij}},
\end{equation}
where $\alpha=(\alpha_{ij})\in M_n(\Z)^+$. This generalizes to
non-invertible transformations the usual action of {\em monomial
Cremona transformations}, which corresponds to the case of
$\alpha \in \GL_n(\Z)$.

\smallskip

As in \eqref{eq:1} above, one considers the distinguished point
$\gamma=(1,1,\ldots,1)$ in $\bT^n$. The endomotive in this case
is constructed by considering the inverse images
$$ X_\alpha = \{ t \in \bT^n \, |\, \alpha(t) = \gamma \} $$
for $\alpha \in M_n(\Z)^+$. These form a projective system, under
composition of the semigroup transformations, and one takes
the corresponding projective limit
\begin{equation}\label{Xprojlimalpha}
 X = \varprojlim_\alpha X_\alpha, 
\end{equation} 
or equivalently the direct limit
$$ A =\varinjlim_\alpha A_\alpha, $$
where $X_\alpha ={\rm Spec}(A_\alpha)$ and $X={\rm Spec}(A)$.
It is shown in \cite{Mar} that this limit is given by 
$$ A \simeq \Q[\Q/\Z]^{\otimes n}. $$
The algebraic endomotive is then given by the $\Q$-algebra 
$\Q[\Q/\Z]^{\otimes n}\rtimes M_n(\Z)^+$ and the analytic endomotive
by the $C^*$-algebra $C^*(\Q/\Z)^{\otimes n}\rtimes M_n(\Z)^+$.
A time evolution on this algebra is obtained by setting
\begin{equation}\label{Tnsigmat}
\sigma_t(\mu_\alpha)=\det(\alpha)^{it} \mu_\alpha \ \ \text{ and } \ \  \sigma_t(a)=a,
\end{equation}
for all $\alpha \in M_n(\Z)^+$ and for all $a\in C^*(\Q/\Z)^{\otimes n}$.

\smallskip

In the construction of \cite{Mar} the choice of the semigroup $S_n =M_n(\Z)^+$
is large, hence one needs to either pass to a convolution algebra that eliminates
the ${\rm SL}_n(\Z)$-symmetry of the spectral levels of the Hamiltonian, or else
computing the partition function with respect to a type II$_1$-trace, which gives
$$ Z(\beta)=\sum_{\alpha\in M_n(\Z)^+/{\rm SL}_n(\Z)} \det(\alpha)^{-\beta}=
\prod_{k=0}^{n-1} \zeta(\beta -k). $$

\smallskip

In the more general construction we present in the rest of this section,
this problem of large ${\rm SL}_n(\Z)$-symmetries
is avoided, by working with the smaller semigroup $S$ that we defined in Section \ref{Ssec}
above. In the case of a single torus, our $S$ in fact reduces to the much smaller
$S=\{ m I\,|\, m\in \N \}$ with $I$ the $n\times n$-identity matrix, which computes the same
projective limit \eqref{Xprojlimalpha} as the larger $S_n$.
The construction of the time evolution we provide for arbitrary toric varieties 
will also be more general than \eqref{Tnsigmat}, hence providing a broader 
range of possible zeta functions.

\medskip
\subsection{Construction of the endomotive}

Let $S$ be the semigroup of an abstract toric variety, constructed in \S \ref{Ssec} above.
For each $\phi$ in $S$, let $X_{\phi}$ denote the preimage under $\phi$ of the set
of distinguished points in all toric orbits,
\begin{equation}\label{Xphi}
X_\phi = \phi^{-1} ( \underset{k=1}{\overset{m}{\sqcup}} \{ \gamma_k \}).
\end{equation}

\smallskip

For any $\phi_{1},\phi_{2}$ and $\varphi$ in $S$ with $\phi_{1}=\phi_{2}\circ \varphi$,
we have a transition map $\varphi:X_{\phi_{1}}\rightarrow X_{\phi_{2}}$, 
given by composition. Thus, all the sets $X_{\phi}$'s together with these maps form a projective
system. We denote by $X$ the projective limit. We have the following result.

\begin{prop}\label{Lemma2.1}
As a topological space, the projective limit is 
\begin{equation}\label{eq:3}
X=\varprojlim_\phi
X_{\phi}\simeq
\underset{k=1}{\overset{m}{\sqcup}}
 {\rm Hom}_{\mathbb{Z}}(\sigma_{k}^{\bot}\cap M,\mathbb{\hat{Z}})
\end{equation}
and the algebra of continuos functions is correspondingly given by
\begin{equation}\label{eq:4}
C(X)\simeq  \bigoplus_{k=1}^m
C^{*}((\sigma_{k}^{\bot}\cap M)\otimes(\mathbb{Q}/\mathbb{Z}))  \, .
\end{equation}
\end{prop}

\proof
Note that any matrix in $S$ is invertible
in $M_{d}(\mathbb{Q})$ and the inverse is also compatible with $\Sigma$
and bijective as a map of cones, thus some integer multiple of this
inverse lies in $M_{d}(\mathbb{Z})$ and hence in $S$, which shows
that for any $\phi_{1}\in S$ there exists $\phi_{2}\in S$ and $n\in\mathbb{N}^{+}$such
that $\phi_{1}\phi_{2}=nI$ (here $I$ denotes the $d$ by $d$ identity
matrix). Therefore $\{nI|n\in\mathbb{N}^{+}\}$ is a cofinal subset
of the indexing set, and it suffices to compute the projective limit
over this subset.

\smallskip

Let $X_{n,k}=(nI)^{-1}(\gamma_{k})$ be the preimage of
the distinguished point $\gamma_{k}$ under the action of $nI$. 
Then it is clear that $$X_{nI}=\underset{k=1}{\overset{m}{\sqcup}}X_{n,k}.$$
Note that each transition map $tI$ preserves the second index $k$
of $X_{n,k}$, {\em i.e.} the image of $X_{tn,k}$ lies in $X_{n,k}$. Thus, 
the projective system $\{X_{nI}|n\in\mathbb{N}^{+}\}$ splits as the
disjoint union of projective systems $\{X_{n,k}|n\in\mathbb{N}^{+}\}$
($k=1,...,m$). Hence we have 
\begin{equation}
X\simeq \varprojlim_n
X_{n}\simeq\underset{k=1}{\overset{m}{\sqcup}}\varprojlim_n X_{n,k}.
\end{equation}

\smallskip

On the other hand, we have natural isomorphisms 
\begin{equation}\label{projXnk}
X_{n,k}=(nI)^{-1}(\gamma_{k})\simeq {\rm Hom}_{\mathbb{Z}}(\sigma_{k}^{\bot}\cap M,\mathbb{Z}/n\mathbb{Z}),
\end{equation}
which give 
\begin{equation}
\varprojlim_n X_{n,k}\simeq {\rm Hom}_{\mathbb{Z}}(\sigma_{k}^{\bot}\cap M,
\varprojlim_n (\mathbb{Z}/n\mathbb{Z}))\simeq {\rm Hom}_{\mathbb{Z}}
(\sigma_{k}^{\bot}\cap M,\mathbb{\hat{Z}}),
\end{equation}
from which the first statement follows immediately.

\smallskip

As for the second one, clearly we have 
\begin{equation}
C(X)\simeq C(\underset{k=1}{\overset{m}{\sqcup}}{\rm Hom}_{\mathbb{Z}}(\sigma_{k}^{\bot}\cap M,\mathbb{\hat{Z}}))\simeq 
\bigoplus_{k=1}^m
C({\rm Hom}_{\mathbb{Z}}(\sigma_{k}^{\bot}\cap M,\mathbb{\hat{Z}})).
\end{equation}
Moreover, by Pontryagin duality, we obtain canonical isomorphisms 
\begin{equation}
C({\rm Hom}_{\mathbb{Z}}(\sigma_{k}^{\bot}\cap M,\mathbb{\hat{Z}}))\simeq C^{*}({\rm Hom}_{\mathbb{Z}}(\sigma_{k}^{\bot}\cap M,\mathbb{\hat{Z}})^{\wedge}),
\end{equation}
where 
\begin{equation}
{\rm Hom}_{\mathbb{Z}}(\sigma_{k}^{\bot}\cap M,\mathbb{\hat{Z}})^{\wedge}\simeq {\rm Hom}_{\mathbb{Z}}({\rm Hom}_{\mathbb{Z}}(\sigma_{k}^{\bot}\cap M,\mathbb{\hat{Z}}),\mathbb{Q}/\mathbb{Z})
\end{equation}
is the Pontryagin dual. Note that $\mathbb{Q}/\mathbb{Z}$ and $\hat{\mathbb{Z}}$
are Pontryagin duals to each other, hence we have a natural isomorphism
\begin{equation}\label{eq11}
{\rm Hom}_{\mathbb{Z}}({\rm Hom}_{\mathbb{Z}}(\sigma_{k}^{\bot}\cap M,\mathbb{\hat{Z}}),\mathbb{Q}/\mathbb{Z})\simeq {\rm Hom}_{\mathbb{Z}}({\rm Hom}_{\mathbb{Z}}(\sigma_{k}^{\bot}\cap M,{\rm Hom}(\mathbb{Q}/\mathbb{Z},\mathbb{Q}/\mathbb{Z})),\mathbb{Q}/\mathbb{Z}).
\end{equation}
Furthermore, 
\begin{equation}
{\rm Hom}_{\mathbb{Z}}(\sigma_{k}^{\bot}\cap M,{\rm Hom}(\mathbb{Q}/\mathbb{Z},\mathbb{Q}/\mathbb{Z}))\simeq {\rm Hom}_{\mathbb{Z}}((\sigma_{k}^{\bot}\cap M)\otimes_{\mathbb{Z}}(\mathbb{Q}/\mathbb{Z}),\mathbb{Q}/\mathbb{Z}),
\end{equation}
showing that we can identify the right-hand-side of \eqref{eq11}
with the double dual of $(\sigma_{k}^{\bot}\cap M)\otimes(\mathbb{Q}/\mathbb{Z})$,
which is canonically isomorphic to $(\sigma_{k}^{\bot}\cap M)\otimes(\mathbb{Q}/\mathbb{Z})$
itself. 

\smallskip

Combining the identities above we obtain the second statement. 
\endproof

\smallskip

For reasons related to the behavior of the resulting zeta functions, we also
want to consider a variant of the construction of Proposition \ref{Lemma2.1},
where instead of working with the space $X$ of \eqref{eq:3}, we work
with a different space $Y$ given by a {\em product}, instead of a {\em disjoint union},
of the limits $\varprojlim_n X_{n,k}$. Namely, we set
\begin{equation}\label{Yspace}
Y := \prod_{k=1}^n \, \varprojlim_n X_{n,k}.
\end{equation}
The algebra of functions is correspondingly given by the {\em tensor
product} instead of {\em direct sum},
\begin{equation}\label{CYalg}
C(Y)= \bigotimes_{k=1}^m C(\varprojlim_n X_{n,k}) = \bigotimes_{k=1}^m
C^*((\sigma_k^\perp \cap M)\otimes (\Q/\Z)),
\end{equation}
where the last equality follows as in Proposition \ref{Lemma2.1}.

\medskip
\subsection{The crossed product algebra and Hilbert space representations}

Note that, since by construction $M$ comes endowed with a basis, we may identify
$M$ as $\mathbb{Z}^{d}$, hence we can write $M\otimes(\mathbb{Q}/\mathbb{Z})$
as $(\mathbb{Q}/\mathbb{Z})^{d}$. In this way, each element of $(\sigma_{k}^{\bot}\cap M)\otimes(\mathbb{Q}/\mathbb{Z})$ can be regarded as a $d$-tuple in $(\mathbb{Q}/\mathbb{Z})^{d}$, and
there is an induced action of $S$ on 
$M\otimes(\mathbb{Q}/\mathbb{Z})\simeq(\mathbb{Q}/\mathbb{Z})^{d}$
given by matrix multiplication, which corresponds to the action
given by the map $\phi\otimes1$ on $M\otimes(\mathbb{Q}/\mathbb{Z})$.

\smallskip

It follows that this action via matrix multiplication again maps each
$(\sigma_{k}^{\bot}\cap M)\otimes(\mathbb{Q}/\mathbb{Z})\subseteq(\mathbb{Q}/\mathbb{Z})^{d}$
into the corresponding $(\sigma_{\phi(k)}^{\bot}\cap M)\otimes(\mathbb{Q}/\mathbb{Z})\subseteq(\mathbb{Q}/\mathbb{Z})^{d}$.
We will make use of this action and this identification later in Lemma
\ref{Lemma2.3} and \ref{Lemma2.4}. 

\smallskip

The $C^*$-algebras of functions 
\begin{equation}\label{abalg}
 C(X)\simeq \bigoplus_{k=1}^m C^*((\sigma_{k}^{\bot}\cap M)\otimes(\mathbb{Q}/\mathbb{Z})) 
\end{equation} 
and
\begin{equation}\label{abalgtens}
 C(Y)\simeq \bigotimes_{k=1}^m C^*((\sigma_{k}^{\bot}\cap M)\otimes(\mathbb{Q}/\mathbb{Z})) 
\end{equation} 
will serve as abelian parts of two variants of the endomotive construction. 
We represent them on Hilbert spaces as follows.

\begin{lem}\label{Lemma2.2}
There is a natural representation of
the algebra $C(X)$ of \eqref{abalg} as bounded linear operators on the Hilbert space 
\begin{equation}\label{HilbSp}
\mathcal{H}_X=\bigoplus_{k=1}^m \ell^{2}(
{\rm Hom}_{\mathbb{Z}}(\sigma_{k}^{\bot}\cap M,\mathbb{Z}))
\end{equation}
and of the algebra $C(Y)$ of \eqref{abalgtens} on
\begin{equation}\label{HilbSptens}
\mathcal{H}_Y=\bigotimes_{k=1}^m \ell^{2}(
{\rm Hom}_{\mathbb{Z}}(\sigma_{k}^{\bot}\cap M,\mathbb{Z})).
\end{equation}
\end{lem}

\proof First we observe that 
\begin{equation}\label{Homcones}
{\rm Hom}_{\mathbb{Z}}((\sigma_{k}^{\bot}\cap M)\otimes(\mathbb{Q}/\mathbb{Z}),\mathbb{Q}/\mathbb{Z})\simeq {\rm Hom}_{\mathbb{Z}}(\sigma_{k}^{\bot}\cap M,{\rm Hom}_{\mathbb{Z}}(\mathbb{Q}/\mathbb{Z},\mathbb{Q}/\mathbb{Z}))\simeq {\rm Hom}_{\mathbb{Z}}(\sigma_{k}^{\bot}\cap M,\mathbb{\hat{Z}}).
\end{equation}
Moreover, $\mathbb{Z}$ naturally embeds into $\mathbb{\hat{Z}}$, hence 
we have a natural embedding of 
\begin{equation}\label{HomZhatZ}
{\rm Hom}_{\mathbb{Z}}(\sigma_{k}^{\bot}\cap M,\mathbb{Z}) \hookrightarrow 
{\rm Hom}_{\mathbb{Z}}(\sigma_{k}^{\bot}\cap M,\mathbb{\hat{Z}}), 
\end{equation}
which means that we can naturally identify each element of 
${\rm Hom}_{\mathbb{Z}}(\sigma_{k}^{\bot}\cap M,\mathbb{Z})$
with a homomorphism 
\begin{equation}\label{fkhom}
f_{k}:(\sigma_{k}^{\bot}\cap M)\otimes(\mathbb{Q}/\mathbb{Z})\rightarrow\mathbb{Q}/\mathbb{Z}.
\end{equation}

\smallskip

We can then represent in the following way 
the algebra $C^{*}((\sigma_{k}^{\bot}\cap M)\otimes(\mathbb{Q}/\mathbb{Z}))$
as bounded linear operators on the Hilbert space $\ell^{2}({\rm Hom}_{\mathbb{Z}}(\sigma_{k}^{\bot}\cap M,\mathbb{Z}))$. Consider the canonical basis 
$$\{\epsilon_{f_{k}}|f_{k}\in {\rm Hom}_{\mathbb{Z}}(\sigma_{k}^{\bot}\cap M,\mathbb{Z})\}, $$
with ${\rm Hom}_{\mathbb{Z}}(\sigma_{k}^{\bot}\cap M,\mathbb{Z})\subset  
{\rm Hom}_{\mathbb{Z}}((\sigma_{k}^{\bot}\cap M)
\otimes(\mathbb{Q}/\mathbb{Z}),\mathbb{Q}/\mathbb{Z})$. Then, 
for each generator 
$e(r_{k})$ of $C^{*}((\sigma_{k}^{\bot}\cap M)\otimes(\mathbb{Q}/\mathbb{Z}))$,
where $r_{k}\in(\sigma_{k}^{\bot}\cap M)\otimes(\mathbb{Q}/\mathbb{Z})$,
we set 
\begin{equation}\label{erfk}
e(r_k)\epsilon_{f_{k}}=\exp(2i\pi f_{k}(r_k))\epsilon_{f_{k}}.
\end{equation}

\smallskip

From now on, for simplicity, we suppress the explicit notation $\exp(2i\pi \cdot)$ and we 
implicitly identify $f_{k}(r)$ with the corresponding root of unity.

\smallskip

Clearly, this preserves the multiplicative structure of the group algebra
since each $f{}_{k}$ is a homomorphism. Thus, we have a naturally
induced representations of the direct sum 
$C(X)\simeq\oplus_{k=1}^m C^{*}((\sigma_{k}^{\bot}\cap M)\otimes(\mathbb{Q}/\mathbb{Z}))$
on $\mathcal{H}_X=\oplus_{k=1}^m
\ell^{2}({\rm Hom}_{\mathbb{Z}}(\sigma_{k}^{\bot}\cap M,\mathbb{Z}))$, 
and of the tensor product 
$C(Y)\simeq\otimes_{k=1}^m C^{*}((\sigma_{k}^{\bot}\cap M)\otimes(\mathbb{Q}/\mathbb{Z}))$
on $\mathcal{H}_Y=\otimes_{k=1}^m
\ell^{2}({\rm Hom}_{\mathbb{Z}}(\sigma_{k}^{\bot}\cap M,\mathbb{Z}))$, respectively given by
\begin{equation}\label{repoplustens}
e(\underline{r})(\oplus_{k=1}^m \epsilon_{f_{k}})=\oplus_{k=1}^m f_{k}(r_{k}) \epsilon_{f_{k}}, \ \ 
\text{ and } \ \  e(\underline{r})(\otimes_{k=1}^m \epsilon_{f_{k}})
=\otimes_{k=1}^m f_{k}(r_{k}) \epsilon_{f_{k}}
\end{equation}
where $\underline{r}=(r_k)_{k=1,\ldots,m}$, and $e(\underline{r}):=\oplus_{k=1}^m e(r_k)$ 
or $e(\underline{r}):=\otimes_{k=1}^m e(r_k)$, respectively.
\endproof

\smallskip
\subsubsection{Semigroup representation}

We represent the semigroup $S$ as bounded on the same Hilbert spaces 
$\mathcal{H}_X$ and $\mathcal{H}_Y$.

\begin{defn}\label{fprimek}
For $f_{k}\in {\rm Hom}(\sigma_{k}^{\bot}\cap M,\mathbb{Z})$, let 
$f'_{1},...,f'_{m}$ denote the compositions $f{}_{1}\circ\phi,...,f{}_{m}\circ\phi$, 
reordered in such a way that $f'_{k}\in {\rm Hom}(\sigma_{k}^{\bot}\cap M,\mathbb{Z})$.
\end{defn}

Notice that it is always possible to reorder the elements
$f{}_{1}\circ\phi,...,f{}_{m}\circ\phi$ as described,
since $\phi$ defines a permutation of the set 
$\{\sigma_{k}^{\bot}\cap M\, | \, k=1,...,m\}$. 
With a slight abuse of notation, in the following we will 
write $\epsilon_{f{}_{k}\circ\phi}$ instead of $\epsilon_{f'_{k}}$. 

\smallskip

\begin{lem}\label{SactH}
The actions given, respectively, by
\begin{equation}\label{muphiplustens}
\mu_{\phi}(\oplus_{k=1}^m \epsilon_{f_{k}})=\oplus_{k=1}^m
\epsilon_{f_{k}\circ \phi} \ \ \text{ and } \ \ \ \mu_{\phi}(\otimes_{k=1}^m \epsilon_{f_{k}})=\otimes_{k=1}^m
\epsilon_{f_{k}\circ \phi},
\end{equation}
define representations of the semigroup $S$ on the Hilbert spaces 
$\mathcal{H}_X$ and $\mathcal{H}_Y$, where in both cases the adjoints are determined by
\begin{equation}\label{mustarsumtens}
\mu_{\phi}^{*} \,\, \epsilon_{f_{k}}= \delta(f_{k}=g_{k}\circ\phi) \, \, \epsilon_{g_{k}}, 
\end{equation}
where
\begin{equation}\label{deltacases}
\delta(f_{k}=g_{k}\circ\phi) = \begin{cases}
1 & f_{k}=g_{k}\circ\phi\\
0 & \mathrm{otherwise.}
\end{cases}
\end{equation}
\end{lem}

\proof By construction, $\phi \mapsto \mu_{\phi}$ 
defines a semigroup homomorphism,  
$$ \mu_{\phi_1 \circ \phi_2} = \mu_{\phi_1} \mu_{\phi_2} , $$
from $S$ to isometries in the 
algebra of bounded operators on $\cH_X$ and $\cH_Y$, respectively. 

The adjoint $\mu^*_{\phi}$ is defined by the relation 
$$ \langle \mu^*_{\phi}\, \xi, \eta \rangle = \langle \xi, 
\mu_{\phi}\, \eta \rangle, $$
for all $\xi,\eta$ in the Hilbert space. It suffices to check the identity on the
elements of the canonical basis, hence we consider $\epsilon_{f_k}$ and
$\epsilon_{h_k}$, for some $f_k$ and $h_k$ in 
${\rm Hom}(\sigma_k^\perp\cap M,\Z)$. Then we have
$$ \langle \xi, \mu_{\phi}\, \eta \rangle =
\delta(h_k = f_k \circ \phi) = 
\langle \mu^*_{\phi}\, \xi, \eta \rangle, $$
for $ \mu^*_{\phi}$ as in \eqref{mustarsumtens}. 
It is clear by construction that $\mu_{\phi}^{*}$ 
is a left inverse of $\mu_{\phi}$ and a right inverse 
on the image of $\mu_{\phi}$.
\endproof

\medskip
\subsubsection{Crossed product relations}

We have the following lemma relating the operators $e(r_{k})$
and $\mu_{\phi}$.

\begin{lem}\label{Lemma2.3}
The action of $S$ on $(\sigma_{k}^{\bot}\cap M)\otimes(\mathbb{Q}/\mathbb{Z})$
satisfies 
\begin{equation}
e(\phi\cdot \underline{r})=\mu_{\phi}^{*}\, e(\underline{r})\, \mu_{\phi}
\end{equation}
\end{lem}

\proof
For any $\epsilon_{f_{k}}$ we have 
$$
\mu_{\phi}^{*}\, e(\underline{r})\, \mu_{\phi}\,\, \epsilon_{f_{k}}=
\mu_{\phi}^{*}e(\underline{r})\,\, \epsilon_{f_{k}\circ\phi} 
 = f_{\phi(k)}\circ\phi(r_{k})\, \,\, \mu_{\phi}^{*}\, \epsilon_{f_{k}\circ\phi}, $$
where by $\phi(k)$ we mean the permutation of
the indices of the $\sigma_{k}^{\bot}\cap M$ induced by $\phi$.
By \eqref{mustarsumtens}, we then write the above as 
$$ f_{\phi(k)}\circ\phi(r_{k})\, \epsilon_{f_k} = e(\phi \cdot \underline{r}) \, 
 \epsilon_{f_{k}}, $$
hence the conclusion holds.
\endproof

Lemma \ref{Lemma2.3} immediately implies the following result.

\begin{lem}\label{Lemma2.4}
Let $\lambda=1$ in the additive case of $(\cA_X,\cH_X)$ and $\lambda=m$
in the multiplicative case $(\cA_Y,\cH_Y)$.
The operators $e(\underline{r})$ and $\mu_{\phi}$
satisfy the relations 
\begin{equation}\label{relmuphi}
\mu_{\phi}\,\, e(\underline{r})\,\, \mu_{\phi}^{*}=
\dfrac{1}{|\det(\phi)|^\lambda}\,\, \underset{\phi\cdot \underline{s}=\underline{r}}{\sum}e(\underline{s}) .
\end{equation}
\end{lem}

\proof
First we show that the number of solutions $\underline{s}$ to the equation 
$\phi\cdot \underline{s}=\underline{r}$ is exactly $|\det(\phi)|$. By definition of the
action of $\phi$ it is clear that $\phi: s_{k'}  \mapsto r_{k}$ only if $s_{k'} 
\in(\sigma_{\phi^{-1}(k)}^{\bot}\cap M)\otimes(\mathbb{Q}/\mathbb{Z})$.

On the other hand, as we observe in Lemma \ref{Lemma2.7} below,
$\sigma_{\phi^{-1}(k)}^{\bot}$ and $\sigma_{k}^{\bot}$ have the same dimension,
and $\phi$ gives an isomorphism between them. Thus, for any $r\in(\mathbb{Q}/\mathbb{Z})^{d}\backslash(\sigma_{\phi^{-1}(k)}^{\bot}\cap M)\otimes(\mathbb{Q}/\mathbb{Z})$,
$(\phi\otimes1)(r)\nsubseteq(\sigma_{k}^{\bot}\cap M)\otimes(\mathbb{Q}/\mathbb{Z})$,
hence, under matrix multiplication, $\phi r\neq r_{k}$. 
Thus, the set of $s_{k'}$ with $\phi\cdot s_{k'}=r_k$ is the set $\{\phi^{-1}r_{k},\phi^{-1}(r_{k}+\underline{1}),...,\phi^{-1}(r_{k}+\underline{(\det(\phi)-1)})\} \mod\mathbb{Z}^{d}$, 
where $\phi^{-1}$ denotes the inverse matrix of $\phi$ in $GL_{d}(\mathbb{Q})$ and
$\underline{1}=(1,1,\ldots, 1)\in \mathbb{Q}^{d}$. This set  can be written as 
 $\phi^{-1} r_k \mod\mathbb{Z}^{d} + {\rm Ker} \phi$,
where ${\rm Ker}\phi$ is a cyclic subgroup of order $\ell =| \det(\phi) |$ of
$(\mathbb{Q}/\mathbb{Z})^{d}$.

\smallskip

Now we prove \eqref{relmuphi}. For each $\underline{s}$ with $\phi\cdot \underline{s}=\underline{r}$,
Lemma \ref{Lemma2.3} gives $e(\underline{r})=\mu_{\phi}^{*}e(\underline{s})\mu_{\phi}$, 
hence 
\begin{equation}\label{mumuphi}
\mu_{\phi}e(\underline{r})\mu_{\phi}^{*}=\mu_{\phi}\mu_{\phi}^{*}\,\,
e(\underline{s})\,\, \mu_{\phi}\mu_{\phi}^{*},
\end{equation}
so that the desired identity holds on the range of $\mu_{\phi}$ on the Hilbert space,
where the projector $\mu_{\phi}\mu_{\phi}^{*}$ is the identity operator.

Now let $\epsilon_{f_k}$ be chosen so that it is not in the range of $\mu_{\phi}$. 
Then, in particular, $f_{\phi^{-1}(k)}\neq f\circ\phi$ for any 
$f\in {\rm Hom}_{\mathbb{Z}}(\sigma_{k}^{\bot}\cap M,\mathbb{Z})$.
The operator on the left-hand-side in \eqref{relmuphi} maps $\epsilon_{f_k}$
to 0, and it suffices to show so does the one on the right-hand-side. 

For simplicity we write 
$\hat r_k$ for $\phi^{-1}r_{k} \mod\mathbb{Z}^{d} \in(\mathbb{Q}/\mathbb{Z})^{d}$.
Then, as noted above, we can write the sum in the right-hand-side of \eqref{relmuphi} as 
a sum over $a_k \in {\rm Ker}\phi$, and the 
action of $e(r_k)$ on $\epsilon_{f_k}$ as multiplication by 
$f_{\phi^{-1}(k)}(\hat r_k+a_k)$. We also reinstate here the explicit notation $\exp(2\pi i\cdot)$
that we suppressed before, so that this action is, in fact, given by multiplication by the phase
factor $\exp(2\pi i\, f_{\phi^{-1}(k)}(\hat r_k))\exp(2\pi i\, f_{\phi^{-1}(k)}(a_k))$. Thus, we can write
the right-hand-side of \eqref{relmuphi} as
\begin{equation}\label{muphirhsplustimes}
\begin{array}{rl}
\displaystyle{\sum_{\phi \cdot s= r_k} e(s) \epsilon_{f_k}}
= & \displaystyle{\sum_{\phi \cdot s=r_k}  f_{\phi^{-1}(k)}(s)\,\, \epsilon_{f_k}}
\\[3mm]
= & \displaystyle{\sum_{a_k \in {\rm Ker}\phi}  \exp(2\pi i\, f_{\phi^{-1}(k)}(\hat r_k))
\exp(2\pi i\, f_{\phi^{-1}(k)}(a_k)) \,\, \epsilon_{f_k}}. 
\end{array}
\end{equation}

Note then that the map $f_{\phi^{-1}(k)}:(\sigma_{\phi^{-1}(k)}^{\bot}\cap M)\otimes(\mathbb{Q}/\mathbb{Z})\rightarrow(\mathbb{Q}/\mathbb{Z})$
cannot be identically zero on ${\rm Ker}\phi$, otherwise it would factor through
$(\sigma_{\phi(k)}^{\bot}\cap M)\otimes(\mathbb{Q}/\mathbb{Z})$,
contradicting the fact that no $f\in Hom_{\mathbb{Z}}(\sigma_{k}^{\bot}\cap M,\mathbb{Z})$ in $Hom_{\mathbb{Z}}((\sigma_{\phi(k)}^{\bot}\cap M)\otimes(\mathbb{Q}/\mathbb{Z}),(\mathbb{Q}/\mathbb{Z}))$
satisfies $f_{\phi^{-1}(k)}=f\circ\phi$. In particular, $\xi:=\exp(2\pi if_{\phi^{-1}(k)}(a))$
is a nontrivial root of unity of order $\ell$, where $a$ is a generator
of the finite cyclic group ${\rm Ker}\phi$ of order $\ell$. Therefore we
have 
\begin{equation}
\underset{a \in {\rm Ker}\phi}{\sum}\exp(2\pi i\,f_{\phi^{-1}(k)}(a))
=\underset{j=1}{\overset{\ell}{\sum}}\xi^{j}=0,
\end{equation}
which in conjunction with \eqref{muphirhsplustimes} shows that
\begin{equation}
\sum_{\phi \cdot \underline{s}=\underline{r}} e(\underline{s})=0 
\end{equation}
on the complement of the range of $\mu_{\phi}$, 
and this completes the proof.
\endproof

\medskip

Now we are ready to define our endomotives for toric
varieties.

\begin{defn}\label{Definition2.5}
The {\em additive} endomotive associated with
a toric variety $X_{\Sigma}$ is given by the semigroup crossed
product algebra 
\begin{equation}\label{Acrossplus}
\mathcal{A}_{X,\Sigma}=C(X)\rtimes_{\rho} S
\simeq (\oplus_{k=1}^m C^{*}((\sigma_{k}^{\bot}\cap M)\otimes(\mathbb{Q}/\mathbb{Z}))
\rtimes_{\rho}S
\end{equation}
and the {\em multiplicative} endomotive associated with $X_\Sigma$ is given by the crossed
product algebra
\begin{equation}\label{Acrosstimes}
\mathcal{A}_{Y,\Sigma}=C(Y)\rtimes_{\rho} S
\simeq (\otimes_{k=1}^m C^{*}((\sigma_{k}^{\bot}\cap M)\otimes(\mathbb{Q}/\mathbb{Z}))
\rtimes_{\rho}S,
\end{equation}
where in both cases the semigroup action is given by 
$\rho_{\phi}(e(\underline{r}))=\mu_{\phi}e(\underline{r})\mu_{\phi}^{*}$
under the representation specified in Lemma \ref{Lemma2.2}.
\end{defn}

Note that, by Lemma \ref{Lemma2.4}, $\rho_{\phi}(e(\underline{r}))=
\mu_{\phi}e(\underline{r})\mu_{\phi}^{*}$ 
is indeed an element of $\oplus_{k=1}^m C^{*}({\rm Hom}_{\mathbb{Z}}(\sigma_{k}^{\bot}\cap M,\mathbb{Q}/\mathbb{Z}))$
(respectively, $\otimes_{k=1}^m C^{*}({\rm Hom}_{\mathbb{Z}}(\sigma_{k}^{\bot}\cap M,\mathbb{Q}/\mathbb{Z})))$, hence $\rho$ gives a well-defined action.

\medskip
\subsection{Time evolution and Hamiltonian}

It is natural to consider on the algebras $\mathcal{A}_{X,\Sigma}$
and $\cA_{Y,\Sigma}$ a time evolution
and quantum statistical mechanical properties that generalize the
corresponding ones of the Bost-Connes system.  
We discuss first the case of the multiplicative endomotive
\eqref{Acrosstimes}, and then the similar case for the additive \eqref{Acrossplus}.

\smallskip

The following result
describes a general approach to construct a time evolution together with
a generating Hamiltonian.

\smallskip

\begin{prop}\label{Lemma2.6}
Let $g:S\rightarrow\mathbb{R}^*_{+}$ be a semigroup
homomorphism, and for $k=1,\ldots, m$, let 
\begin{equation}\label{hk}
h_k: {\rm Hom}_{\mathbb{Z}}(\sigma_{k}^{\bot}\cap M,\mathbb{Z})\rightarrow\mathbb{R}^*_{+}
\end{equation}
be positive-valued functions such that
\begin{equation}\label{hkghk}
h_k(f'_{k})=g(\phi)\, h_k(f_{k}), 
\end{equation}
for all $f_{k}\in {\rm Hom}_{\mathbb{Z}}(\sigma_{k}^{\bot}\cap M,\mathbb{Z})$ and for
all $\phi\in S$, where the $f'_{k}$ are as in Definition \ref{fprimek}. Let $\lambda=1$
in the additive case of $(\cA_X,\cH_X)$ and $\lambda=m$ in the multiplicative
case of $(\cA_Y, \cH_Y)$. Then setting
\begin{equation}\label{sigmatgk}
\sigma_{t}(\mu_{\phi})=g(\phi)^{i\lambda t}\, \mu_{\phi},  \ \ \  \text{ and } \ \ \
\sigma_{t}(e(\underline{r}))=e(\underline{r}),
\end{equation}
for all $\phi\in S$ and for all $r_k\in (\sigma_{k}^{\bot}\cap M)\otimes(\Q/\mathbb{Z})$,
with $\underline{r}=(r_{k})$, determines a time evolution 
$\sigma_t$ on the endomotive algebras $\mathcal{A}_X$ and $\cA_Y$, generated by
a Hamitonian $H$ determined by the operator
\begin{equation}\label{Hamghk}
H\,\,  \epsilon_{f_{k}}=\log(h_k(f_{k}))\,\, \epsilon_{f_{k}}, 
\end{equation} 
for all $f_{k}\in {\rm Hom}_{\mathbb{Z}}(\sigma_{k}^{\bot}\cap M,\mathbb{Z})$.
\end{prop}

\proof As in Lemma \ref{Lemma2.2}, with a slight 
abuse of notation, we write $h_k(f_{k}\circ\phi)$ instead of $h_k(f'_{k})$.

It is easy to check that \eqref{sigmatgk} defines a
one-parameter family of automorphisms, both in the
case of $\cA_X$ and of $\cA_Y$. The operator 
\eqref{Hamgh} uniquely determines self-adjoint positive 
linear operators, which we still denote by $H$, on 
both $\mathcal{H}_X$ and $\cH_Y$. We check that $H$ is indeed
the Hamiltonian. In both the additive and the multiplicative case
we have 
\begin{equation}\label{erH}
e^{itH}e(\underline{r})e^{-itH}=e(\underline{r})e^{itH}e^{-itH}=e(\underline{r}).
\end{equation}
In the additive case of $\cA_X$ and $\cH_X$ we have
\begin{equation}\label{musigmahg}
\begin{array}{rl}
e^{itH}\mu_{\phi}e^{-itH}\,\, \underset{k=1}{\overset{m}{\oplus}}\epsilon_{f_{k}}= &
\underset{k=1}{\overset{m}{\oplus}}h_k(f_{k})^{-it}\, h_k(f_{k}\circ\phi)^{it}\,\, \epsilon_{f_{k}\circ \phi} 
\\[3mm]
= & g(\phi)^{it}\,\, \underset{k=1}{\overset{m}{\oplus}}\epsilon_{f_{k}\circ \phi} \\[3mm]
= & \sigma_{t}(\mu_{\phi})\,\, \underset{k=1}{\overset{m}{\otimes}}\epsilon_{f_{k}}.
\end{array}
\end{equation}
Similarly, in the multiplicative case of $\cA_Y$ and $\cH_Y$, we have
\begin{equation}\label{musigmahg2}
\begin{array}{rl}
e^{itH}\mu_{\phi}e^{-itH}\,\, \underset{k=1}{\overset{m}{\otimes}}\epsilon_{f_{k}}= &
\underset{k=1}{\overset{m}{\otimes}}h_k(f_{k})^{-it}\, h_k(f_{k}\circ\phi)^{it}\,\, \epsilon_{f_{k}\circ \phi}  
\\[3mm]
= & g(\phi)^{imt}\,\, \underset{k=1}{\overset{m}{\otimes}}\epsilon_{f_{k}\circ \phi} \\[3mm]
= & \sigma_{t}(\mu_{\phi})\,\, \underset{k=1}{\overset{m}{\otimes}}\epsilon_{f_{k}}.
\end{array}
\end{equation}
This proves the statement.
\endproof

In the multiplicative case, we can consider the more general form of
the time evolution given below, which agrees with the one above in
the case where $h=h_1\otimes\cdots\otimes h_m$.

\begin{cor}\label{L2.6cor}
Let $g:S\rightarrow\mathbb{R}^*_{+}$ be a semigroup
homomorphism, and 
$$ h:\underset{k=1}{\overset{m}{\prod}}{\rm Hom}_{\mathbb{Z}}(\sigma_{k}^{\bot}\cap M,\mathbb{Z})\rightarrow\mathbb{R}^*_{+} $$
be a positive-valued function such that 
\begin{equation}\label{hgh}
h(f'_{1},...,f'_{m})=g(\phi)\, h(f_{1},...,f_{m}),   \forall f_{k}\in {\rm Hom}_{\mathbb{Z}}(\sigma_{k}^{\bot}\cap M,\mathbb{Z}),\forall\phi\in S
\end{equation}
for all $ f_{k}\in {\rm Hom}_{\mathbb{Z}}(\sigma_{k}^{\bot}\cap M,\mathbb{Z})$ and for all $\phi\in S$.
Then setting 
\begin{equation}\label{sigmatg}
\sigma_{t}(\mu_{\phi})=g(\phi)^{it}\, \mu_{\phi},\  \forall\phi\in S, \ \  \text{ and } \ \ 
\sigma_{t}(e(\underline{r}))=e(\underline{r}),\ \forall \underline{r}=(r_{k})
\in \prod_{k=1}^m (\sigma_{k}^{\bot}\cap M)\otimes(\Q/\mathbb{Z}),
\end{equation}
defines a time evolution $\sigma:\mathbb{R}\rightarrow Aut(\mathcal{A}_Y)$
generated by the Hamitonian 
\begin{equation}\label{Hamgh}
H\,\, (\otimes_{k=1}^m \epsilon_{f_{k}})=\log(h(f_{1},...,f_{k}))\, \underset{k=1}{\overset{m}{\otimes}}\epsilon_{f_{k}}, \ \ \forall f_{k}\in {\rm Hom}_{\mathbb{Z}}(\sigma_{k}^{\bot}\cap M,\mathbb{Z}).
\end{equation}
\end{cor}

\proof As before, we write $h(f_{1}\circ\phi,...,f_{m}\circ\phi)$ instead of $h(f'_{1},...,f'_{m})$.
The argument follows exactly as in Proposition \ref{Lemma2.6}, with \eqref{musigmahg2}
replaced by
\begin{equation}\label{musigmahg3}
\begin{array}{rl}
e^{itH}\mu_{\otimes,\phi}e^{-itH}(\underset{k=1}{\overset{m}{\otimes}}\epsilon_{f_{k}})= &
h(f_{1},...,f_{k})^{-it}\, h(f_{1}\circ\phi,...,f_{m}\circ\phi)^{it}\, (\underset{k=1}{\overset{m}{\otimes}}\epsilon_{f_{k}\circ \phi}) \\[3mm]
= & g(\phi)^{it}\, (\underset{k=1}{\overset{m}{\otimes}}\epsilon_{f_{k}\circ \phi}) \\[3mm]
= & \sigma_{t}(\mu_{\phi})(\underset{k=1}{\overset{m}{\otimes}}\epsilon_{f_{k}}).
\end{array}
\end{equation}
\endproof

\medskip
\subsubsection{Symmetries}

By symmetries of a $C^*$-dynamical system $(\cA,\sigma_t)$ we mean here a group
$\cG\subset {\rm Aut}(\cA)$ of automorphisms of the $C^*$-algebra of observables
that is compatible with the time evolution, $\sigma_t \circ \gamma = \gamma \circ \sigma_t$,
for all $t\in \R$ and for all $\gamma\in \cG$. 

\smallskip

\begin{lem}\label{auto}
The group $\cG={\rm Hom}_\Z(\sigma^\perp_k\cap M, \hat\Z^*)$ acts on
$(\cA_{X,\Sigma},\sigma_t)$ and $(\cA_{Y,\Sigma},\sigma_t)$
as symmetries.
\end{lem}

\proof For $\gamma\in \cG$ we set $\gamma\, \mu_\phi =\mu_\phi$, for all
$\phi\in S$, and on $C^*((\sigma^\perp_k\cap M)\otimes \Q/\Z)$ as
$\gamma(v\otimes t)=v\otimes \gamma_v(t)$,
for $v\otimes t$ in $(\sigma^\perp_k\cap M)\otimes \Q/\Z$ and
$\gamma\in \Hom_\Z(\sigma^\perp_k\cap M,\hat\Z^*)$, with $\gamma_v=\gamma(v)\in \hat\Z^*$.
The action is compatible with the time evolution, since $\sigma_t$ fixes the
algebra $C^*((\sigma^\perp_k\cap M)\otimes \Q/\Z)$ and only acts nontrivially on
the semigroup elements $\mu_\phi$, which are fixed by $\cG$.
\endproof

\smallskip

It was shown in \cite{CoMa} that often, in the setting of quantum statistical mechanical 
systems it is useful to consider not only symmetries given by automorphisms but also
endomorphisms, but here we will restrict our attention to automorphisms,
as that will suffice for our purposes.

\medskip
\subsection{The partition function}

One obtains from the covariant representations of Proposition \ref{Lemma2.6}
the following zeta functions.

\begin{lem}\label{Zetabetas}
The partition functions of the systems described in Proposition \ref{Lemma2.6}
are of the form
\begin{equation}\label{Zsum}
Z(\beta)=\sum_{k=1}^m \sum_{f\in {\rm Hom}_\Z(\sigma_k^\perp\cap M, \Z)} h_k(f)^{-\beta},
\end{equation}
in the additive case of $(\cA_X,\sigma_t)$ represented on $\cH_X$, and
\begin{equation}\label{Zprod}
Z(\beta)=\prod_{k=1}^m \sum_{f\in {\rm Hom}_\Z(\sigma_k^\perp\cap M, \Z)} h_k(f)^{-\beta}
\end{equation}
in the multiplicative case $(\cA_Y,\sigma_t)$ on $\cH_Y$, which in the
case of Corollary \ref{L2.6cor} takes the form
\begin{equation}\label{Zprodh}
Z(\beta)=\sum_{\underline{f}=(f_k) \in \prod_k {\rm Hom}_\Z(\sigma_k^\perp\cap M, \Z)}
h(f_{1},...,f_{m})^{-\beta}.
\end{equation}
\end{lem}

\proof
The partition function for the system $(\cA_X,\sigma_t)$ with the
covariant representation on $\cH_X$ with Hamiltonian $H$ as in
Proposition \ref{Lemma2.6} is given by
\begin{equation}\label{Zsum1}
Z(\beta)={\rm Tr}(e^{-\beta H})=
\sum_{\oplus_{k=1}^m \epsilon_{f_{k}}\in \cH_X}  \langle \oplus_{k=1}^m \epsilon_{f_{k}},
e^{-\beta H}\,\,  \oplus_{k=1}^m \epsilon_{f_{k}} \rangle =\sum_{f_k\in \sqcup_{k=1}^m
{\rm Hom}_\Z(\sigma_k^\perp\cap M, \Z)} h_k(f_k)^{-\beta}.
\end{equation}
The partition function for the system $(\cA_Y, \sigma_t)$, with the time
evolution specified in Proposition \ref{Lemma2.6}, is given by 
\begin{equation}\label{Ztensor1}
Z(\beta)={\rm Tr}(e^{-\beta\, H})=
\underset{\underset{k=1}{\overset{m}{\otimes}}\epsilon_{f_{k}}\in\mathcal{H}_Y}{\sum}
\langle \underset{k=1}{\overset{m}{\otimes}}\epsilon_{f_{k}}, e^{-\beta H} \underset{k=1}{\overset{m}{\otimes}}\epsilon_{f_{k}}\rangle =
\sum_{\underline{f}=(f_k) \in \prod_k {\rm Hom}_\Z(\sigma_k^\perp\cap M, \Z)}
(h_1(f_{1}) \cdots h_m(f_{m}))^{-\beta},
\end{equation}
which clearly extends to the case of Corollary \ref{L2.6cor} as
\begin{equation}\label{Ztensor}
Z(\beta )= 
\underset{\underset{k=1}{\overset{m}{\otimes}}\epsilon_{f_{k}}\in\mathcal{H}_Y}{\sum}
\langle \underset{k=1}{\overset{m}{\otimes}}\epsilon_{f_{k}}, e^{-\beta H} \underset{k=1}{\overset{m}{\otimes}}\epsilon_{f_{k}}\rangle =
\sum_{\underline{f}=(f_k) \in \prod_k {\rm Hom}_\Z(\sigma_k^\perp\cap M, \Z)}
h(f_{1},...,f_{m})^{-\beta}.
\end{equation}
\endproof

\smallskip

\begin{rem}\label{sumvsprod}{\rm
The main difference between the additive and the multiplicative cases
$\cA_X$ and $\cA_Y$, for a quantum statistical mechanical perspective
is that, while the additive case corresponds to the geometric decomposition
of the toric variety into torus orbits, the multiplicative case corresponds,
as a quantum mechanical system, to regarding the torus orbits as
independent systems, hence the decomposition of the partition
function as a product of the partition functions associated to the
different orbits.}\end{rem}

\smallskip

In practice one has many ways to define appropriate functions
$g$ and $h$ and thus the corresponding partition functions. Here
we describe one natural way. First we need some additional
results on the semigroup $S$.

\smallskip
\subsubsection{Semigroup and symmetries of the fan}

Recall that $S_{0}$ is the semigroup of all nonsingular
linear transformations of $N$ that is compatible with the fan $\Sigma$
and bijective as a map of the relative interiors of cones in $\Sigma$.
Since the lattice $N$ has a fixed basis, we may identify elements
of $S_{0}$ with $d\times d$ matrices. We have the following lemma
regarding the elements in $S_{0}$.

\begin{lem}\label{Lemma2.7}
Let $\phi\in S_{0}$, then the following holds:
\begin{enumerate}
\item $n\phi\in S_{0}$ for all $n\in\mathbb{N}$;
\item $\phi:\sigma_{k}\rightarrow\sigma_{\phi(k)}$ is a bijective linear
map of cones, for all $k$;
\item $\phi$ leaves every cone invariant if and only if $\phi=nI$, for some $n\in\mathbb{N}$;
\item $\phi^{T}:\sigma_{k}^{\bot}\rightarrow\sigma_{\phi^{-1}(k)}^{\bot}$
is an isomorphism of vector spaces, for all $k$.
\end{enumerate}
\end{lem}

\proof These statements follow immediately from the geometry
of the fan. \endproof

\smallskip

\begin{lem}\label{G0sbgrp}
Let $G_0 \subset S_0$ be defined by
\begin{equation}\label{G0}
G_{0}=\{\phi\in S_{0}\,|\, \phi\neq n\phi', \,\, \forall\phi'\in S_{0},\, \forall n\in \N, \, n>1 \}.
\end{equation}
Then $G_0$ is a subgroup of the permutation group of the cones
of $\Sigma$.
\end{lem}

\proof
Note that, two distinct elements $\phi_{1}\neq \phi_{2}\in G_{0}$ cannot
induce the same permutation of cones in $\Sigma$, 
otherwise we would have an element $\phi_{2}^{-1}\phi_{1}\in \GL_{d}(\mathbb{Q})$
that keeps all the cones invariant, which means that there exists an element
$t\in\mathbb{N}$, such that $t\phi_{2}^{-1}\phi_{1}\in S_{0}$, and therefore 
$t\phi_{2}^{-1}\phi_{1}=nI$, for some $n\in\mathbb{N}$, leading to a contradiction. 
Thus, we can identify $G_0$ with a subgroup of the permutation group of the cones
in $\Sigma$. 
\endproof

\smallskip

\begin{rem}\label{detG0}{\rm
Although it is not true in general that $G_{0}$ is a
subgroup of $\GL_{d}(\mathbb{Z})$, i.e. that every element in $G_{0}$
has determinant $\pm1$, it does hold in many cases, including the
case of the fan corresponding to the projective space $\mathbb{P}^{d}$. 
}\end{rem}

\smallskip
\subsubsection{Time evolution and $G$-orbits}

Now we restrict our attention to those toric varieties $X_\Sigma$ for which
the group $G_{0}$ of \eqref{G0} is indeed a subgroup of $\GL_{d}(\mathbb{Z})$.
In these cases we have $S_{0}=\mathbb{N} \times G_{0}$. Correspondingly
we have $S=\mathbb{N} \times G$, where $G$ is the group consisting
of transposes of elements in $G_{0}$, which is again a subgroup of
$\GL_{d}(\mathbb{Z})$. 

\smallskip

Let $N_{k}$ is the sublattice of $N$ consisting of vectors 
orthogonal to $\sigma_{k}^{\bot}\cap M$, or equivalently 
the intersection of $N$ with the vector space spanned
by $\sigma_{k}$. 

\smallskip

\begin{thm}\label{ghbetaG}
Let $\Sigma$ be a fan for which $S=\mathbb{N} \times G$, with
$G\subset \GL_d(\Z)$. Given the choice of a basis of $N/N_1 \simeq \Z^{d_1}$, let
$N_1^*=\{ \xi \in N/N_1\,|\, \gcd(\xi)=1\}$, where $\gcd(\xi)$
is the greatest common divisor of the coordinates of $\xi \in \Z^{d_1}$.
The choice of a constant $c\in {\mathbb R}$ and of a function $h_1: N_1^* \to {\mathbb R}^*_+$
that is constant on the orbits of the subgroup $G_1\subset G$ 
that fixes $\sigma_1$ determine functions $g$ and $h$ satisfying the
condition \eqref{hgh}, which therefore define a time evolution on the (multiplicative) 
endomotive of the toric variety.
\end{thm}

\proof By duality we can naturally identify each 
${\rm Hom}_{\mathbb{Z}}(\sigma_{k}^{\bot}\cap M,\mathbb{Z})$
with the quotient $N/N_{k}$. If $f_{k}$ corresponds to $\xi_{k}$ under this
identification, then $f_{k}\circ\phi$ corresponds to $\phi^{T}\xi_{k}$. 
Note that, since $\phi$ defines a permutation of the cones, there is also
an action of $G$ on the set of cones by permutation. 

\smallskip

We define $g$ as follows. Fix a constant $c\in\mathbb{R}$. Each element
in $S$ can be uniquely written as $n\phi$ for some $n\in\mathbb{N}$ 
and some $\phi\in G$. Then we set 
\begin{equation}\label{gnc}
g(n\phi)=n^{c}. 
\end{equation}

\smallskip

As for $h$, we define it separately for each orbit of the action of $G$ on the set of cones.
Suppose $\{\sigma_{1},...,\sigma_{p}\}$ is such an orbit, and let
$G_{1}$ be the subgroup of $G$ that leaves $\sigma_{1}$ invariant.
Assuming for the moment that we have defined a positive-valued map
$h_{1}$ on $N/N_{1}$ that is compatible with the action of $\mathbb{N} \times G_{0}$.
We define $h_{j}$ on $N/N_{j}$ by fixing a $\phi\in G$ that takes
$\sigma_{1}$ to $\sigma_{j}$ and setting 
\begin{equation}
h_{j}(\phi^{T}\xi)=h_{1}(\xi),\ \   \forall\xi\in N/N_{1}.
\end{equation}
It is easy to check that this is well-defined since $\phi^{T}$ gives
an isomorphism between $N/N_{1}$ and $N/N_{j}$ and $h_{1}$ is compatible
with the action of $\mathbb{N}\times G_{0}$. In this way, we define
positive valued functions $h_{1},...,h_{p}$ that are together compatible
with the action of $S$, and we have, for each $j$,
\begin{equation}
\underset{\xi_{j}\in N/N_{j}}{\sum}h_{j}(\xi_{j})^{-s}=\underset{\xi_{1}\in N/N_{1}}{\sum}h_{1}(\xi_{1})^{-s}.
\end{equation}

\smallskip

To define $h_{1}$, we choose a basis of $N/N_{1}$, and we identify
it with vectors in $\mathbb{Z}^{d_{1}}$, where $d_{1}$ is the rank
of the quotient lattice $N/N_{1}$. Note that this is indeed a lattice, 
since it is isomorphic to ${\rm Hom}_{\mathbb{Z}}(\sigma_{k}^{\bot}\cap M,\mathbb{Z})$.
Then consider the orbit of the action of $G_{1}$ on $N/N_{1}$. For
each $\xi\in\mathbb{Z}^{d_{1}}$, let $\gcd(\xi)$ denote, as above, the greatest
common divisor of the $d_{1}$ coordinates of $\xi$. Then it can
be easily shown that $\gcd(\xi)$ is the same for all $\xi$ in
one orbit. Let $N_{1}^{*}=\{\xi\in N/N_{1}\simeq\mathbb{Z}^{d_{1}}\, | \,\gcd(\xi)=1\}$.
Then it follows that $G_{1}$ also acts on $N_{1}^{*}$. Let $h{}_{1}:N_{1}^{*}\rightarrow
\mathbb{R}^*_{+}$
be any function that is constant on each orbit, and we extend $h{}_{1}$
to $N/N_{1}$ by setting 
\begin{equation}
h{}_{1}(n\xi^{*})=n^{c}h{}_{1}(\xi^{*}),\forall\xi^{*}\in N_{1}^{*},
\end{equation}
which gives the desired $h_{1}$.
Finally we set 
\begin{equation}\label{hfkhbeta}
 h(f_{1},...,f_{m})=\overset{m}{\underset{k=1}{\prod}}h_{k}(\xi_{k}).
\end{equation} 
Then it is clear that the desired identity 
$$ h(f{}_{1}\circ\phi,...,f{}_{m}\circ\phi)=g(\phi)\, h(f_{1},...,f_{m}) $$
holds for these functions.
\endproof

In this case the partition function is then of the following form. 

\begin{cor}\label{Zcor}
Let $\Sigma$ be a fan for which $S=\mathbb{N} \times G$, with
$G\subset \GL_d(\Z)$, and let $\sigma_t$ be a time-evolution on the endomotives
of the toric variety, defined using the functions $g$ and $h$ constructed 
as in Theorem \ref{ghbetaG}.
The partition function is then of the form
\begin{equation}\label{ZetaG}
Z(\beta)=\zeta(\beta)^{m}\underset{k=1}{\overset{m}{\prod}}(\underset{\xi_{k}^{*}\in N_{k}^{*}}{\sum}h_{k}(\xi_{k}^{*})^{-\beta}).
\end{equation}
\end{cor}

\proof We have
$$ Z(\beta)=\underset{(\xi_{1},...,\xi_{m})\in\underset{k=1}{\overset{m}{\prod}}N/N_{k}}{\sum}\underset{k=1}{\overset{m}{\prod}}h_{k}(\xi_{k})^{-\beta}, $$
which we write equivalently as
$$Z(\beta)= \underset{k=1}{\overset{m}{\prod}}(\underset{\xi_{k}\in N/N_{k}}{\sum}h_{k}(\xi_{k})^{-\beta})=\zeta(\beta)^{m}\underset{k=1}{\overset{m}{\prod}}(\underset{\xi_{k}^{*}\in N_{k}^{*}}{\sum}h_{k}(\xi_{k}^{*})^{-\beta}). $$
\endproof

\begin{rem}\label{remfan} {\rm
In the most general cases, if the fan $\Sigma$ is quite large and
lacks symmetry, then the semigroup $S$ only consists of matrices
of the form $nI$ (alternativesly, for any fan $\Sigma$ we can always
replace $S$ in our above construction with the sub-semigroup $\{nI|n\in\mathbb{N}^{+}\}$
and everything else follows exactly the same). In this case $G$ is
the trivial group and thus we can define each $h_{k}$ on $N/N_{k}$
independently, and $h_{k}$ can be obtained from an arbitrary positive-valued
function on $N_{k}^{*}$. }
\end{rem}

A more concrete example of the construction of Theorem \ref{ghbetaG}
is obtained in the following way.

\begin{cor}\label{normh}
Let $\| \cdot \|_{k}$ be a norm on each real vector space $(N/N_{k})_{\mathbb{R}}$.
Setting $h_{k}(\xi)=\|\xi \|_{k}^{c}$ satisfies
the identity  \eqref{hgh}, with $g$ as in \eqref{gnc}, hence it defines
a time evolution. The corresponding partition function on the (multiplicative)
endomotive is given by 
\begin{equation}\label{normZ}
Z(\beta)=\underset{k=1}{\overset{m}{\prod}}(\underset{\xi\in N/N_{k}}{\sum}||\xi||_{k}^{-c\beta}).
\end{equation}
\end{cor}

The additive cases are analogous, with the partition functions given by sums instead
of products over the set of torus orbits.

\smallskip
\subsubsection{Projective spaces}

We consider the example of projective spaces,
where several of the general properties discussed above can be
seen more explicitly.

\smallskip

\begin{lem}\label{GMPd}
In the case of $\mathbb{P}^{d}$, the orbit space of $M$ under 
the action of $G$ is the orbit space of the subspace of 
$\mathbb{Z}^{d+1}$ given by solutions of $x_{1}+...+x_{d+1}=0$ 
under coordinate permutations.
\end{lem}

\proof
The fan $\Sigma_{\Delta_{d}}$ associated to $\mathbb{P}^{d}$,
consists of the cones generated by all proper subsets of $\{e_{0},e_{1},...,e_{d}\}$, 
where $e_{1},...,e_{d}$ form the standard basis of $N$ and $e_{0}=-e_{1}-...-e_{d}$.
Then it is not hard to see that $G$ consists of all matrices whose
$d$ rows are $d$ distinct vectors from $\{e_{0},e_{1},...,e_{d}\}$.
Then the orbit of a point $(a_{1},...,a_{d})$ in $M$ consists of
all vectors whose $d$ coordinates are $d$ different elements of
the multiset $\{a_{1},a_{2},...,a_{d},-a_{1}-...-a_{d}\}$. Therefore, 
we may identify the orbit space of $M$ under $G$ with the set of
all $(d+1)$-subsets of $\mathbb{Z}$ that sum up to 0, or equivalently
the orbit space of the subspace of $\mathbb{Z}^{d+1}$ defined by
$x_{1}+...+x_{d+1}=0$ under the action of coordinate permutation.
\endproof

\smallskip

The corresponding partition functions implicitly encode the information
about this symmetry. In the Section \ref{HeightSec} below 
we will give a more concrete construction for projective spaces, 
using the arithmetic height functions.

\medskip

\section{Endomotives of toric varieties and $\F_1$-geometry}\label{F1sec}

There are currently many different approaches aimed at developing a form
of algebraic geometry over the ``field with one element" $\F_1$. For an overview
of various contribution and their interrelatedness, we refer to reader to the
survey \cite{LLP2}. 

\smallskip

Toric varieties play a crucial role in $\F_1$. They are the only class of $\Z$-varieties
that admit $\F_1$-structures according to {\em all} of the existing variants of $\F_1$-geometry.
In some of the strongest formulations, they are essentially the only varieties that
descend to $\F_1$ (see for instance \cite{Vezz} for a comparative
analysis). While other approaches allow for a broader range of varieties over $\F_1$,
toric varieties remain an important class on which different constructions can be compared.

\smallskip

We discuss here the relation between the endomotive construction
and the $\F_1$-structure on the toric variety. Relations between endomotives
and $\F_1$-geometry had already been considered in \cite{CCM2}, for the case of
the Bost--Connes endomotive, and in \cite{Mar}, for its multivariable generalizations.
We will consider here three different connections to $\F_1$-geometry: the relation
between the semigroup action and the $\Lambda$-ring structure, following Borger's
approach to $\F_1$-geometry via $\Lambda$-rings, \cite{Bor}; the relation to Soul\'e's
notion of varieties over $\F_1$, as in \cite{Soule}; a weaker form of the endomotive
construction that extends from the case of toric variety to $\F_1$-varieties defined
by torified spaces, in the sense of \cite{LLP1}.

\medskip
\subsection{Endomotives and $\Lambda$-ring structures}

As in \cite{Bor}, an integral $\Lambda$-ring structure on a commutative ring $R$, whose
underlying abelian group is torsion free, is given by an action of the semigroup
$\N$ by endomorphisms of $R$, so that, for each prime $p$, the action $\phi_p$
of $p$ on $R$ is a Frobenius lift, 
\begin{equation}\label{Froblift}
\phi_p(r) - r^p \in pR, \ \ \forall r\in R.
\end{equation}
As in \cite{Mar}, a $\Q$-algebra $A$ has a $\Lambda$-ring structure if it has an
action of $\N$ by endomorphisms, and $A=R\otimes \Q$, with $R$ a commutative
ring with a $\Lambda$-ring structure as above, inducing the same $\N$-action on $A$.

\smallskip

In order to compare the construction of endomotives of toric varieties given above
with $\Lambda$-ring structures, we need to work with algebraic endomotives
instead of the analytic ones discussed above.

\smallskip

\begin{defn}\label{algendotoric}
Given an abstract toric variety $X_\Sigma$ defined over $\Q$, 
the algebraic abelian subalgebra of the endomotives of $X_\Sigma$
are, respectively, the $\Q$-algebras 
\begin{equation}\label{Qendos}
\oplus_{k=1}^m \Q[(\sigma^\perp_k \cap M)\otimes \Q/\Z], \ \ \ \text{ and } \ \ \
\otimes_{k=1}^m \Q[(\sigma^\perp_k \cap M)\otimes \Q/\Z].
\end{equation}
The semigroup $S$ acts on both by endomorphisms, and one obtains the
algebraic additive and multiplicative endomotives of $X_\Sigma$ as the
algebraic semigroup crossed products
\begin{equation}\label{QendosS}
\cA_{X,\Sigma,\Q}= (\oplus_{k=1}^m \Q[(\sigma^\perp_k \cap M)\otimes \Q/\Z])\rtimes S, 
\ \ \ \text{ and } \ \ \
\cA_{Y,\Sigma,\Q}=(\otimes_{k=1}^m \Q[(\sigma^\perp_k \cap M)\otimes \Q/\Z])\rtimes S.
\end{equation}
We also consider the subalgebras $\cA_{X,\Sigma\Q}^0$ and $\cA_{Y,\Sigma\Q}^0$
obtained, as above, as algebraic crossed products by the subsemigroup $S_0\subset S$,
with $S_0=\{ n I\, | \, n\in \N \}$.
\end{defn}

\smallskip

\begin{prop}\label{endoLambda}
Let $X_\Sigma$ be an abstract toric variety defined over $\Q$.
The abelian algebras \eqref{Qendos} are direct limits of $\Lambda$-rings, with
the $\Lambda$-ring structure given by the action of $S_0=\{ n I\, | \, n\in \N \}$.
In the additive case, there are embeddings
\begin{equation}\label{Xnkembed}
\sqcup_{k=1}^m X_{n,k}\simeq \sqcup_{k=1}^m
{\rm Hom}_\Z(\sigma_k^\perp\cap M,\Z/n\Z) \hookrightarrow \sqcup_{k=1}^m O(\sigma_k) 
\subseteq X_\Sigma,
\end{equation} 
determined by embeddings of ${\rm Hom}_\Z(\sigma_k^\perp\cap M,\Z/n\Z)$
into the torus orbit $O(\sigma_k)$ of $X_\Sigma$, which induce corresponding maps of
$\Lambda$-rings, with respect to the $\Lambda$-ring structure on the toric variety.
\end{prop}

\proof 
The abelian algebra $\oplus_{k=1}^m \Q[(\sigma^\perp_k \cap M)\otimes \Q/\Z]$ is a direct limits of
$A_n=\oplus_{k=1}^m \Q[(\sigma^\perp_k \cap M)\otimes \Z/n\Z]=R_n\otimes \Q$ with
$R_n=\oplus_{k=1}^m \Z[(\sigma^\perp_k \cap M)\otimes \Z/n\Z]$. The action of $\N$ 
is the one given by $e(r_k) \mapsto e(\phi_p(r_k))$, which on the limit $A=\varinjlim_n A_n$
corresponds to $\mu_\phi^* e(r_k) \mu_\phi$. 

The embeddings \eqref{Xnkembed} are determined by the identification 
$O(\sigma_{k})\simeq {\rm Hom}_{\mathbb{Z}}(\sigma_{k}^{\bot}\cap M,\bG_m)$, 
as in \eqref{maporbs}. In particular, the subsemigroup $S_0$ acts
on the part of the limit set $X_k=\varprojlim_n X_{n,k}$ in the torus 
$T_{\sigma_k^\perp\cap M}=O(\sigma_{k})$ as a restriction to 
$X_k \subset T_{\sigma_k^\perp\cap M}$ of the action of $\N$ on the torus 
$T_{\sigma_k^\perp\cap M}$ given on the coordinates by $\sigma_p: t_j \mapsto t_j^p$.
The compatibility of this action with the Frobenius action, showing that it defines a
$\Lambda$-ring structure, follows in the same way as the analogous result 
for the original Bost--Connes case, given in \cite{CCM2} and \cite{Mar}. 
The embeddings \eqref{Xnkembed} determine maps of 
$\Lambda$-rings, since the $\Lambda$-ring structure on a toric variety $X_\Sigma$ is
compatible with the decomposition into torus orbits, namely, it induces 
the compatible $\Lambda$-ring structures described above
on all the torus orbits $O(\sigma_{k})$, see \S 2.4 of \cite{Bor}.
\endproof

\medskip
\subsection{Endomotives and Soul\'e's varieties over $\F_1$}

A relation between the endomotive of the Bost--Connes system and Soul\'e's notion
of varieties over $\F_1$ was described in \cite{CCM2}, based on the construction of
a model over $\Z$ of the endomotive. We show here that, in a similar way, we can
obtain models over $\Z$ for the endomotives of toric varieties.

\smallskip
\subsubsection{Integer models of the endomotives}

As in the Bost--Connes  case analyzed in \cite{CCM2}, the crossed product 
algebras \eqref{QendosS}
of our algebraic endomotives of toric varieties admits a model over $\Z$, which is obtained
in the following way. 

\smallskip

Consider the algebras $\cA_{X,\Sigma,\Z}$ and $\cA_{Y,\Sigma,\Z}$ generated by
$$ \cC_{X,\Sigma,\Z}:=\oplus_{k=1}^m \Z[(\sigma^\perp_k \cap M)\otimes \Q/\Z] $$ 
and 
$$ \cC_{Y,\Sigma,\Z}:= \otimes_{k=1}^m \Z[(\sigma^\perp_k \cap M)\otimes \Q/\Z], $$
respectively, and by elements $\mu_\phi^*$ and $\tilde\mu_\phi$, for all $\phi\in S$,
satisfying the relations $\tilde\mu_{\phi_1}\tilde\mu_{\phi_2}=\tilde\mu_{\phi_1\phi_2}$,
$\mu^*_{\phi_1}\mu^*_{\phi_2}=\mu^*_{\phi_1\phi_2}$, for all $\phi_1,\phi_2\in S$, and
$\mu_{\phi}^* \tilde\mu_\phi = |\det(\phi)|^\lambda$,
where $\lambda=1$ in the additive case and $\lambda=m$ in the multiplicative case, and
$$ \mu_\phi^*\, a = \sigma_\phi(a)\, \mu_\phi^*  \ \ \ \text{ and } \ \ \
 a\, \tilde\mu_\phi = \tilde\mu_\phi \, \sigma_\phi(a), $$
for all $\phi \in S$ and for all $a\in \cC_{X,\Sigma,\Z}$ or $\cC_{Y,\Sigma,\Z}$,
where $\sigma_\phi(e(\underline{r}))=e(\phi \cdot \underline{r})$.

\smallskip
\subsubsection{Soul\'e's gadgets and varieties}

Soul\'e's approach to $\F_1$-geometry is based on the concept of
an $\F_1$-gadget and of an $\F_1$-variety. A gadget consists of data $(X,\cA_X,e_{x,\sigma})$
with $X: \cR \to {\rm Sets}$ a covariant functor from the category $\cR$ of finitely
generated flat rings, $\cA_X$ a complex algebra, and evaluation maps given
by algebra homomorphisms $e_{x,\sigma}: \cA_X \to \C$, for all $x\in X(R)$ and
$\sigma: R\to \C$, satisfying $e_{f(y),\sigma}=e_{y,\sigma\circ f}$, for any ring
homomorphism $f: R'\to R$. An affine variety $V_\Z$ over $\Z$ determines a gadget
with $X_V(R)={\rm Hom}(\cO(V),R)$ and $\cA_X=\cO(V)\otimes \C$. A gadget is an
affine $\F_1$-variety is $X(R)$ is finite and there is an affine variety $W_\Z$ with 
a morphism of gadgets $X \to X_W$ such that morphisms of gadgets 
$X \to X_V$ are induced by morphisms of varieties $W_\Z\to V_\Z$.
Heuristically, one should think of the case where $R=\Z[\Z/n\Z]$, for which $X(R)$
gives the cyclotomic points.

\smallskip
\subsubsection{Endomotives as $\F_1$-varieties}

In \cite{CCM2}, the Bost--Connes endomotive is described in terms of a family of
$\F_1$-varieties $\mu^{(k)}$, in the sense of Soul\'e, determined by the functor
$\underline{\mu}^{(k)}: \cR \to {\rm Sets}$ given by $\underline{\mu}^{(k)}(R)=\{ r\in R\,|\, r^k=1\}$,
represented by $\underline{\mu}^{(k)}(R)={\rm Hom}_\Z(\Z[\Z/k\Z],R)$. The algebra $\cA_{\mu^{(k)}}$
of the gadget is given by $\Z[\Z/k\Z]\otimes_\Z \C=\C[\Z/k\Z]$. The projective limit
$\mu^{(\infty)}=\varprojlim_k \mu^{(k)}$ is given by the functor that assigns 
$\underline{\mu}^{(\infty)}(R)={\rm Hom}_\Z(\Z[\Q/\Z],R)$.
The maps in this projective limit are exactly the ones used in the
construction of the Bost--Connes endomotive, coming from the action of the semigroup $\N$.
This tower of zero-dimensional affine $\F_1$-varieties $\mu^{(k)}$ describes the inductive system of
extensions $\F_{1^k}$, as defined by Kapranov--Smirnov, with $\F_{1^\infty}=\varinjlim_k \F_{1^k}$,
where the ``extension of coefficients to $\Z$" is formally given by  
$$\F_{1^k}\otimes_{\F_1}\Z := \Z[t,t^{-1}]/(t^k-1),$$ 
with $\Z[t, t^{-1}]/(t^k-1)\simeq \Z[\Z/k\Z]$.

\smallskip

We have an analogous construction in the case of endomotives of toric varieties.

\begin{prop}\label{endoF1}
Let $X_\Sigma$ be an abstract toric variety defined over $\Z$. 
To the abelian parts $\cC_{X,\Sigma,\Z}$ and  $\cC_{Y,\Sigma,\Z}$ 
of the endomotives $\cA_{X,\Sigma,\Z}$ and $\cA_{Y,\Sigma,\Z}$ 
one can assign projective systems of affine
$\F_1$-varieties in the sense of Soul\'e, where the maps in the
projective systems are induced by the action of the semigroup $S$ of the endomotives.
\end{prop}

\proof Consider the functors $\underline{\mu}_{X,\Sigma}^{(n)}: \cR \to {\rm Sets}$
given by 
$$ \underline{\mu}_{X,\Sigma}^{(n)}(R)={\rm Hom}_\Z(\oplus_{k=1}^m 
\Z[(\sigma_k^\perp\cap M)\otimes \Z/n\Z], R) $$
$$ \underline{\mu}_{Y,\Sigma}^{(n)}(R)={\rm Hom}_\Z(\otimes_{k=1}^m 
\Z[(\sigma_k^\perp\cap M)\otimes \Z/n\Z], R) . $$
Using $X_{n,k}\simeq {\rm Hom}_\Z(\sigma_k^\perp\cap M, \Z/n\Z)$ and
${\rm Hom}_\Z(\sigma_k^\perp\cap M,{\rm Hom}_\Z(\Q/\Z,\Q/\Z))\simeq 
{\rm Hom}_\Z({\rm Hom}_\Z((\sigma_k^\perp\cap M)\otimes_\Z \Q/\Z, \Q/\Z)$,
we see that the varieties $\mu_{X,\Sigma}^{(n)}$ associated to the above
functors form a projective system where the projective limits are,
respectively, $\cC_{X,\Sigma,\Z}$ and  $\cC_{Y,\Sigma,\Z}$, with the maps
of the projective system coming from the elements $\phi\in S$. The analytic
datum of the $\F_1$-gadgets is given by the $\C$-algebras
$\cC_{X,\Sigma,\Z}\otimes_\Z \C$ and  $\cC_{Y,\Sigma,\Z}\otimes_\Z \C$.
The construction is otherwise completely analogous to the case discussed in \cite{CCM2}.
\endproof

\smallskip

Unlike the $\Lambda$-ring structure of the toric variety discussed above, which replies
of semigroup actions providing consistent liftings of the Frobenius action, the construction
described here uses cyclotomic points on the toric variety and it 
fits with Soul\'e's and Manin's general philosophy, \cite{Man}, \cite{Soule} of 
cyclotomy as descent data from $\Z$ to $\F_1$.

\medskip
\subsection{Endomotives and torified spaces}

We discuss here another approach to $\F_1$-geometry, based on ``torifications",
which was developed by L\'opez-Pe\~{n}a and Lorscheid in \cite{LLP1}. We recall their
main definition of a torified space.

\begin{defn}\label{torified}
Let $X$ be a variety over $\Z$. A {\em torification} of $X$ is a disjoint union $T=\sqcup_{j\in I}T_j$
of tori $T_j=\bG_m^{d_j}$, together with a morphism $e_X: T \to X$, such that $e_X|_{T_j}$ 
is an immersion for all $j$ and $e_X$ a bijection of the set of $\bK$-points, 
$T(\bK) \simeq X(\bK)$, over any field $\bK$.
\end{defn}

A toric variety is a torified space, through its decomposition into torus orbits 
$X_\Sigma=\sqcup_{k=1}^m O(\sigma_k)$. However,
the notion of torification is much more general and it includes, for 
example, spaces with cell decompositions. More restrictive conditions on the torification (affine,
regular) can be imposed that restrict the class of (affinely, regularly) torified spaces, 
see \cite{LLP2} for more details. For our purposes, we do not impose any of these stronger conditions,
and we consider torifications as in Definition \ref{torified} above. We show that 
a simple variant of the construction of the multivariable Bost--Connes endomotives of
\cite{Mar} provides endomotives associated to arbitrary torified spaces, which generalize
(in a weaker form) the construction we described for toric varieties. 

\begin{prop}\label{torifiedendo}
Let $X$ be a variety over $\Z$, which admits a torification, as in Definition \ref{torified}, and
let $T=\sqcup_{j\in I}T_j$ with $T_j=\bG_m^{d_j}$ be a choice of a torification on $X$. For each
torus $T_j$ of the torification, consider the projective system
$$ X_n(T_j)=\{ t\in \bT^{d_j}\,|\, s_n(t)=\gamma \}, $$
with $\gamma=(1,1\ldots,1)\in \bT^{d_j}$ and with $s_n: t=(t_i)_{i=1,\ldots,d_j} \mapsto
s_n(t)=(t_i^n)_{i=1,\ldots,d_j}$. The semigroup $\N$ acts by endomorphisms on the
projective limit $X(T_j)=\varprojlim_n X_n(T_j)$ and on the algebra $\Q[\Q/\Z]^{\otimes d_j}$
with ${\rm Spec}(X(T_j))=\Q[\Q/\Z]^{\otimes d_j}$. This determines an additive algebraic 
endomotive $\cA_{X(T),\Q}:=(\oplus_{j\in I} \Q[\Q/\Z]^{\otimes d_j})\rtimes \N$ and a
multiplicative $\cA_{Y(T),\Q}:=(\otimes_{j\in I} \Q[\Q/\Z]^{\otimes d_j})\rtimes \N$. The
corresponding analytic endomotives are given by $\cA_{X(T)}:=(\oplus_{j\in I} 
C^*(\Q/\Z)^{\otimes d_j})\rtimes \N$ and $\cA_{Y(T)}:=(\otimes_{j\in I} 
C^*(\Q/\Z)^{\otimes d_j})\rtimes \N$, respectively. There are representation by
bounded operators of $\cA_{X(T)}$ and $\cA_{Y(T)}$ on the Hilbert spaces
$\cH_{X(T)}=\oplus_{j\in I} \ell^2(\N^{d_j})$ and 
$\cH_{Y(T)}=\otimes_{j\in I}\ell^2(\N^{d_j})$, respectively. 
Suppose given a semigroup homomorphism $g:\N \to \R^*_+$ and $\R^*_+$-valued
functions $h_j$ on $\N^{d_j}$ satisfying $h_j(s_n(\underline{n}_j))=g(n)\, h_j(\underline{n}_j)$,
for all $\underline{n}_j\in \N^{d_j}$ and $n\in \N$ with $s_n(\underline{n}_j)=n \underline{n}_j$
the coordinatewise multiplication. Then setting $\sigma_t(\mu_n)=g(n)^{it}$ and 
$\sigma_t(e_j(\underline{r}))=e_j(\underline{r})$, for all $n\in \N$, for all $e_j(\underline{r})$ in 
$C^*(\Q/\Z)^{\otimes d_j})$, for all $j\in i$ and all $t\in \R$, defines a time evolution on 
$\cA_{X(T)}$ and $\cA_{Y(T)}$, respectively, whose Hamiltonian is determined by the operator
$$ H \epsilon_{\underline{n}_j}= \log ( h_j(\underline{n}_j) )\,\, \epsilon_{\underline{n}_j}, $$
where $\{ \epsilon_{\underline{n}_j} \}$ denotes the canonical orthonormal basis of $\ell^2(\N^{d_j})$.
\end{prop}

\proof Everything follows the same construction as in \cite{Mar}, using only 
the subsemigroup $\{ n I_{d_j} \}$ of $M_{d_j}(\Z)^+$ for the endomotive
construction, so we will not reproduce the details here. The construction of
the time evolution and Hamiltonian is modeled on the analogous construction
we gave in Section \ref{EndoSec} for toric varieties and the argument follows in the same way.
\endproof

\smallskip

While in the endomotive construction for toric varieties the semigroup $S$ 
depends not only on the decomposition of $X_\Sigma$ into tori, but also on
how these tori fit together as orbits of the same torus action (through the 
$G$ subgroup of $S$), the construction for torified spaces is necessarily
weaker and only contains the information on the decomposition into tori
given by a choice of torification. One can think of the action of the semigroup $\N$
in the endomotive of a torified space that is not a toric variety as a weaker replacement
for a $\Lambda$-ring structure, associated to a choice of torification.

\section{Height functions and endomotives}\label{HeightSec}

Now we generalize the construction described in Section \ref{EndoSec} to
build endomotives of projective toric varieties, and study their properties
using the arithmetic height function. The arithmetic height function is
a key notion in diophantine geometry and it encodes much information
concerning the arithmetic of the varieties, so it seems  particularly interesting
to include this kind of data as part of the quantum statistical mechanics
of the endomotives of toric varieties.

\smallskip

We will focus on the concrete examples of projective spaces and
affine spaces, though much of what we describe can be generalized
to other toric varieties and their height functions.

\smallskip

\medskip
\subsection{Height functions and toric varieties}

Height functions play an important role in addressing questions on the
distribution of rational points on algebraic varieties, see \cite{BaMan}.

\smallskip

For a variety $X$ defined over a number field $\bK$, endowed with a
choice of a line bundle $\cL$ with an adelic metric and a section $s$ in a neighborhood
$U$ of a point $x \in X(\bK)$, one defines a height function as
$$ H_{\bK,\cL,s}(x) = \prod_{v\in {\rm Val}(\bK)} \| s(x) \|^{-1}_v. $$
For an overview of the properties of this type of functions, with respect to
dependence on the data $\bK$, $\cL$, $s$, we refer the reader to the 
survey \cite{ChLoir}. We simply write $H_\cL(x)$ in the following.

\smallskip
\subsubsection{Height zeta functions}
The height zeta function is the associated generating function,
$$ Z_{X,\bK}(\cL,\beta)=\sum_{x\in X(\bK)} H_{\cL}(x)^{-\beta}. $$
For $\cL$ an ample line bundle, Northcott's theorem implies
that $N_X(\cL,B)=\#\{ x\in X(\bK)\,|\, H_\cL(x) \leq B\}$ grows at
most polynomially on $B$. If $N_X(\cL,B)< B^a$, then $Z_{X,\bK}(\cL,\beta)$
converges for $\Re(\beta)>a$, see   \cite{ChLoir} for more details.
For a general overview of properties of the height zeta functions
and applications to the study of algebraic points on varieties, we
also refer the reader to the survey \cite{Man95}. 
For more background on the arithmetic height function, see also \cite{BomGu} and \cite{Sil}.

\smallskip

Height functions on toric varieties and the behavior of the height zeta function
were studied in \cite{BaTsch}, \cite{BaTsch2},
where it is shown that, for $\cL$ with Chern class
in the interior of the cone of effective divisors, the height zeta function on a smooth
projective toric variety gives an asymptotic formula for the number of rational 
points of bounded height of the form $N_X(\cL,B)\sim c(X,\cL,\bK)\, B^{a(\cL)}(\log B)^{b(\cL)-1}$
where the exponents $a(\cL)$ and $b(\cL)$ and the constant $c(X,\cL,\bK)$ are determined
by the geometry of $X$ according to a conjecture of Manin's.

\smallskip
\subsubsection{Heights on projective spaces}
On projective spaces $\bP^d$, one can see that the information
coming from the height function is carried by the archimedean
valuation, with the $p$-adic factor equal to one, see \S 3.1 of \cite{Sil}. 
Thus, one sets 
$$ H_\Q(x)=\max \{ |x_0|_\infty, \ldots, |x_d|_\infty \}, $$
for $x\in \bP^d(\Q)$ and, which extends to points $x\in \bP^d(\bK)$ over 
number fields $\bK$ by
$$ H_{\bK}(x)= H_\bQ(x)^{[\bK:\bQ]}. $$
The {\em absolute height} of a point $x\in \bP^d(\bar\Q)$,
denoted by $H(x)$, is defined as $H(x)=H_\bK(x)^{1/[\bK:\Q]}$, where $\bK$ is a number
field such that $x\in \bP^d(\bar\Q)$, so that the result is independent of $\bK$.
The {\em absolute logarithmic height} on $\bP^d(\bar\Q)$
is the function 
\begin{equation}\label{logheight}
h(x)= \log(H(x)). 
\end{equation}

\smallskip

\begin{rem}\label{logzeta}{\rm
Clearly, an analogous height zeta function for the logarithmic height
would not be convergent, but, as we will see below, one can restrict
to suitable choices of subsets of the set of algebraic points that cut
down the multiplicities to logarithmic size, for which one is then able
to define a zeta function with the desired properties based on the
logarithmic height. This choice seems unnatural from the usual point
of view of height functions in diophantine geometry, but we will see that
it is instead quite natural from the point of view of endomotives. }
\end{rem}

\medskip
\subsection{Endomotives of projective spaces}\label{PdSec}

First we focus on the case of the projective space $\mathbb{P}^{d}$. 
This is a toric variety, with the lattice of one parameter subgroups
given by $N=\mathbb{Z}^{d+1}/(1,...,1)$.

\smallskip

Fix a homogeneous coordinate system on the projective space $\mathbb{P}^{d}$,
then, as we have seen in Lemma \ref{GMPd}, the
elements of the group $G$ are precisely the permutations of coordinates. 

\smallskip
\subsubsection{Algebraic points and the endomotive}

Let $X_{0}\in\mathbb{P}^{d}(\mathbb{\bar{Q}})$ be a finite subset of
the $\mathbb{\bar{Q}}$-algebraic points in the projective space that
is invariant under the action of $G$. We replace the set of distinguished
points in the torus orbits, which we used in our previous construction, with 
the set $X_{0}$, and we consider the preimages of $X_{0}$ under the 
action of the semigroup $S$, which again form a projective system. 

\smallskip

The argument of Proposition \ref{Lemma2.1} applies
exactly in the same way here. It shows that the projective limit $X$
is the disjoint union of the limits corresponding to each individual
point in $X_{0}$. More precisely, as topological spaces, we have
\begin{equation}\label{projlimPd}
X=\varprojlim_\phi \phi^{-1}(X_{0})=\underset{x\in X_{0}}{\bigsqcup}\varprojlim_\phi
\phi^{-1}(x)=\underset{x\in X_{0}}{\bigsqcup}\varprojlim_n \, (nI)^{-1}(x),
\end{equation}
and the corresponding function algebras satisfy
\begin{equation}\label{XoplusPd}
C(X)=\underset{x\in X_{0}}{\bigoplus}C(\varprojlim_n \, (nI)^{-1}(x)).
\end{equation}

\smallskip

As before, we also consider a multiplicative version, where instead
of the disjoint union $X$ we consider the product 
\begin{equation}\label{YlimPd}
Y=\underset{x\in X_{0}}{\prod}\varprojlim_n \, (nI)^{-1}(x),
\end{equation}
and the corresponding algebra
\begin{equation}\label{YoplusPd}
C(Y)=\underset{x\in X_{0}}{\bigotimes}C(\varprojlim_n \, (nI)^{-1}(x)).
\end{equation}

\smallskip

Now we analyze more closely the algebra $C(\varprojlim_n (nI)^{-1}(x))$.
We obtain a more explicit description as follows.

\begin{prop}\label{Lemma3.1}
The algebra $C(\varprojlim_n (nI)^{-1}(x))$
can be identified with 
\begin{equation}\label{algcoords}
C(\varprojlim_n (nI)^{-1}(x)) \simeq 
C^{*}({\rm Hom}_{\mathbb{Z}}(\mathbb{Z}{}^{\ell+1}/(1,...,1),\mathbb{Q}/\mathbb{Z})),
\end{equation}
where $\ell+1$ is the number of nonzero coordinates of $x$.
\end{prop}

\proof Let $x=(x_{0},...,x_{d})$. Without loss of generality, 
we suppose that $x_{0},x_{1}, \ldots ,x_{\ell}$ are all the nonzero coordinates.
Then 
$$ 
(nI)^{-1}(x)=\{(\sqrt[n]{x_{0}},...,\sqrt[n]{x_{\ell}},0,..,0)\in\mathbb{P}^{d}(\mathbb{\mathbb{\bar{Q}}})\}, 
$$
where $\sqrt[n]{x_{j}}$ denotes an $n$-th root of $x_{j}$. We show
that the projective system consisting of the $(nI)^{-1}(x)$
is isomorphic to the natural projective system consisting of the 
$(\mathbb{Z}/n\mathbb{Z})^{\ell+1}/(1,...,1)$, 
as projective systems of discrete topological spaces.

\smallskip

For each $z\in\bar{\mathbb{Q}}^{*}$, we can uniquely write $z=re^{i\theta}$
where $r>0$ and $0\leq\theta<2\pi$. Then we write $z^{1/n}=r^{1/n}e^{\frac{i\theta}{n}}$
for $n\in\mathbb{N}$, where $r^{1/n}$ is the unique $n$-th root
of $r$ in $\mathbb{R}^*_{+}$. Clearly, we have 
$(z^{1/n_{1}n_{2}})^{n_{2}}=z^{1/n_{1}},\forall n_{1},n_{2}\in\mathbb{N}$.
With this notation, each $\sqrt[n]{x_{j}}$ can be uniquely written
as $x_{j}^{1/n}\xi_{n}^{k}$, for some $k\in\{0,1,...,n-1\}$, where
$\xi_{n}=e^{\frac{2\pi i}{n}}$. Therefore, the set of $(d+1)$-tuples
$(\sqrt[n]{x_{0}},...,\sqrt[n]{x_{\ell}},0,..,0)$ can be identified
with $(\mathbb{Z}/n\mathbb{Z})^{\ell+1}$. In terms of homogeneous
coordinates, we have $(nI)^{-1}(x)\simeq(\mathbb{Z}/n\mathbb{Z})^{\ell+1}/(1,...,1)$
as discrete spaces. Clearly we have the following commutative diagram:
\begin{equation}\label{diaglims}
\xymatrix{}
\xymatrix{(n_{1}n_{2}I)^{-1}(x)\ar[d]\ar[r]^{\ n_{2}I} & (n_{1}I)^{-1}(x)\ar[d]\\
(\mathbb{Z}/n_{1}n_{2}\mathbb{Z})^{\ell+1}/(1,...,1)\ar[r]^{\ n_{2}} & (\mathbb{Z}/n_{1}\mathbb{Z})^{\ell+1}/(1,...,1)\ensuremath{}
}
\end{equation}
where the upper map $n_{2}I$ raises the homogeneous coordinates to
$n_2$-th powers and the lower map is multiplication by $n_{2}$. Therefore, 
\begin{equation}\label{limsZell}
\varprojlim_n (nI)^{-1}(x)\simeq \varprojlim_n (\mathbb{Z}/n\mathbb{Z})^{\ell+1}/(1,...,1)\simeq(\mathbb{Z}^{\ell+1}/(1,...,1))\otimes\hat{\mathbb{Z}}
\end{equation}
as topological spaces, which implies isomorphisms of the corresponding commutative $C^*$-algebras
of functions
\begin{equation}\label{limsalg}
C(\varprojlim_n (nI)^{-1}(x))\simeq C((\mathbb{Z}{}^{\ell+1}/(1,...,1))\otimes\hat{\mathbb{Z}}).
\end{equation}
By Pontryagin duality, we also have
\begin{equation}\label{limsdual}
\begin{array}{rl}
C((\mathbb{Z}{}^{\ell+1}/(1,...,1))\otimes\hat{\mathbb{Z}})\simeq &
C^{*}({\rm Hom}_{\mathbb{Z}}((\mathbb{Z}{}^{\ell+1}/(1,...,1))\otimes\hat{\mathbb{Z}},\mathbb{Q}/\mathbb{Z}))
\\[3mm]
\simeq & C^{*}({\rm Hom}_{\mathbb{Z}}(\mathbb{Z}{}^{\ell+1}/(1,...,1),\mathbb{Q}/\mathbb{Z})).
\end{array}
\end{equation}

Note that here by $ $$\mathbb{Z}{}^{\ell+1}/(1,...,1)$ we actually
mean the subgroup $(\mathbb{Z}{}^{\ell+1}\oplus0^{d-\ell})/(1,...,1)$ of
$N=\mathbb{Z}^{d+1}/(1,...,1)$, where  the coordinates that vanish in
$\mathbb{Z}^{d+1}$ correspond to the coordinates of $x$ that vanish.
This embedding into $N$ provides a way to identify the algebras for
points $x$ in different torus orbits of the projective space.
\endproof

\smallskip
\subsubsection{Hilbert space representations}

As in Lemma \ref{Lemma2.2}, we construct crossed product
algebras of the endomotive and represent them on Hilbert spaces.

\smallskip

\begin{lem}\label{Lemma3.2}
Fix a nontorsion element $\alpha\in\bar{\mathbb{Q}^{*}}$.
For $\ell=\ell(x)$ as above, let 
\begin{equation}\label{Balphax}
\mathcal{B}_{\alpha,x}:=\{(\alpha^{k_{0}}x_{0}^{k'_{0}},...,\alpha^{k_{\ell}}x_{\ell}^{k'_{\ell}})\in\mathbb{P}^{d}(\mathbb{\bar{Q}})\,|\, k_{0},k'_{0},\ldots,k_{\ell},k'_{\ell}\in\mathbb{Z}\}.
\end{equation}
Then there is a natural representation of the $C^{*}$-algebra $C^{*}({\rm Hom}_{\mathbb{Z}}(\mathbb{Z}^{\ell+1}/(1,...,1),\mathbb{Q}/\mathbb{Z}))$
associated to $x$ on the Hilbert space $\ell^{2}(\mathcal{B}_{\alpha,x})$,
given by 
\begin{equation}
e(f)\epsilon_{\alpha^{k}x^{k'}}=\exp(2\pi if(k))\, \epsilon_{\alpha^{k}x^{k'}},
\end{equation}
for all $f\in Hom_{\mathbb{Z}}(\mathbb{Z}^{\ell+1}/(1,...,1),\mathbb{Q}/\mathbb{Z})$, and for all 
$k\in\mathbb{Z}{}^{\ell+1}/(1,...,1),k'\in\mathbb{Z}$,
where we abbreviate $(\alpha^{k_{0}}x_{0}^{k'},...,\alpha^{k_{\ell}}x_{\ell}^{k'})$
by $\alpha^{k}x^{k'}$.
\end{lem}

The proof is straghtforward and thus omitted.

\smallskip

\begin{cor}\label{repsBalpha}
It follows that we have naturally induced representations of the algebras
$$ C(X)=\bigoplus_{x\in X_{0}} C(\varprojlim_n (nI)^{-1}(x)) $$
and
$$ C(Y)=\bigotimes_{x\in X_{0}} C(\varprojlim_n (nI)^{-1}(x)), $$
respectively, on the Hilbert spaces
\begin{equation}\label{l2Bsum}
 \bigoplus_{x\in X_{0}} \ell^2(\mathcal{B}_{\alpha,x})=\ell^2(\mathcal{B}_{X,\alpha}), 
\ \ \ \text{ where } \ \  \mathcal{B}_{X,\alpha}=\underset{x\in X_{0}}{\bigsqcup}\mathcal{B}_{\alpha,x},
\end{equation}  
and
\begin{equation}\label{l2Bprod}
 \bigotimes_{x\in X_{0}} \ell^2(\mathcal{B}_{\alpha,x})=\ell^2(\mathcal{B}_{Y,\alpha}), 
\ \ \ \text{ where } \ \  \mathcal{B}_{Y,\alpha}=\underset{x\in X_{0}}{\prod}\mathcal{B}_{\alpha,x}.
\end{equation} 
\end{cor}

Again, this follows immediately from the previous statement.
We then represent the semigroup on the same Hilbert spaces in the following way.

\begin{lem}\label{SrepX0}
The semigroup $S$ has representations on the Hilbert spaces \eqref{l2Bsum} and
\eqref{l2Bprod}, determined by setting
\begin{equation}\label{phiBalpha}
\mu_{\phi} \epsilon_{\alpha^{k}x^{k'}}= \epsilon_{\phi\cdot(\alpha^{k}x^{k'})},\ \ \ 
\forall\phi\in S,\  \forall \alpha^{k}x^{k'} \in\mathcal{B}_{\alpha,x},\ \forall x\in X_{0}.
\end{equation}
\end{lem}

\proof
Here by $ $$\phi\cdot(\alpha^{k}x^{k'})$ we mean the action of $S$
on $\mathbb{P}^{d}(\bar{\mathbb{Q}})$. Note that we can write $\phi=n\phi_{0}$
for some $n\in\mathbb{N}$ and $\phi_{0}\in G$, then 
\begin{equation}\label{phiBaction}
\phi\cdot(\alpha^{k}x^{k'})=\phi_{0}\cdot(\alpha^{nk}x^{nk'})=\alpha^{nk}(\phi_{0}\cdot x)^{nk'}\in\mathcal{B}_{\alpha,\phi_{0}\cdot x}
\end{equation}
where by $\phi_{0}\cdot x$ we mean the action of $G$ on $X_{0}$.
It is then clear that \eqref{phiBalpha} has the right properties and defines a
representation of the semigroup $S$ by isometries of the Hilbert space.
\endproof

Moreover, just like in the previous section, the adjoint of the isometry $\mu_{\phi}$
is given by 
\begin{equation}
\mu_{\phi}^{*}\epsilon_{\alpha^{k}x^{k'}}=\begin{cases}
\epsilon_{\alpha^{r}x^{r'}} & \alpha^{k}x^{k'} =\phi\cdot(\alpha^{r}x^{r'})\\
0 & \mathrm{otherwise.}
\end{cases}
\end{equation}

As in Lemma \ref{Lemma2.3} and Lemma \ref{Lemma2.4}, 
the following lemma relates the operators $e(f)$ and $\mu_{\phi}$.

Let  $\phi\cdot f_x$ denote the action of $S$ on 
${\rm Hom}_{\mathbb{Z}}(\mathbb{Z}{}^{\ell(x)+1}/(1,...,1),\mathbb{Q}/\mathbb{Z})$
given by precomposition by $\phi$, considered as a linear map on
the lattice $N=\mathbb{Z}^{d+1}/(1,...,1)$.

\begin{lem}\label{Lemma3.3}
We have the following identities
\begin{equation}\label{efmurel1}
e(\phi\cdot f_{x})=\mu_{\phi}^{*}e(f_{x})\mu_{\phi}
\end{equation}
and
\begin{equation}\label{efmurel2}
\mu_{\phi}e(f_{x})\mu_{\phi}^{*}= \frac{1}{n}  \underset{\phi\cdot f=f_{x}}{\sum}e(f)
\end{equation}
for all $\phi\in S$, and all 
$f_{x}\in {\rm Hom}_{\mathbb{Z}}(\mathbb{Z}{}^{\ell(x)+1}/(1,...,1),\mathbb{Q}/\mathbb{Z})$, 
for all $x\in X_{0}$. 
\end{lem}

\proof
The proof is completely analogous to those of Lemma \ref{Lemma2.2} and 
Lemma \ref{Lemma2.3} and are therefore omitted. The action $\phi\cdot f_x$  on
${\rm Hom}_{\mathbb{Z}}(\mathbb{Z}{}^{\ell(x)+1}/(1,...,1),\mathbb{Q}/\mathbb{Z})$
determines corresponding actions of $S$ on
$$ \underset{x\in X_{0}}{\bigsqcup}
{\rm Hom}_{\mathbb{Z}}(\mathbb{Z}{}^{\ell(x)+1}/(1,...,1),\mathbb{Q}/\mathbb{Z}) $$
and
$$ \underset{x\in X_{0}}{\prod}
{\rm Hom}_{\mathbb{Z}}(\mathbb{Z}{}^{\ell(x)+1}/(1,...,1),\mathbb{Q}/\mathbb{Z}), $$
respectively, satisfying the relations \eqref{efmurel1},  \eqref{efmurel2}, where
$\# \{ f\, |\, \phi\cdot f=f_{x} \}=n$.
\endproof

\smallskip

We have crossed product algebras of the endomotive in the following form.

\begin{defn}\label{endocrossalgpts}
The additive endomotive associated to $\P^d$ with the choice of a finite set $X_0$
of $\mathbb{\bar{Q}}$-algebraic points is the crossed product algebra
\begin{equation}\label{AX0sum}
\begin{array}{rl}
\cA_{X,X_0}:= C(X)\rtimes_\rho S =& (\oplus_{x\in X_0} C(\varprojlim_n (nI)^{-1}(x)))\rtimes_\rho S
\\[3mm]
= & (\oplus_{x\in X_0} C^{*}({\rm Hom}_{\mathbb{Z}}(\mathbb{Z}{}^{\ell(x)+1}/(1,...,1),
\mathbb{Q}/\mathbb{Z})))\rtimes_\rho S
\end{array}
\end{equation}
and the multiplicative endomotive
\begin{equation}\label{AX0prod}
\begin{array}{rl}
\cA_{Y,X_0}:= C(Y)\rtimes_\rho S = & (\otimes_{x\in X_0} C(\varprojlim_n (nI)^{-1}(x)))
\rtimes_\rho S \\[3mm]
= & (\otimes_{x\in X_0} C^{*}({\rm Hom}_{\mathbb{Z}}(\mathbb{Z}{}^{\ell(x)+1}/(1,...,1),
\mathbb{Q}/\mathbb{Z})))\rtimes_\rho S.
\end{array}
\end{equation}
\end{defn}

\medskip
\subsubsection{Time evolution, Hamiltonian, and height function}

We construct time evolutions on the endomotives \eqref{AX0sum} and \eqref{AX0prod}
using the same general technique that we described in the previous section.
We need a preliminary lemma on the behavior of the logarithm height function
under the action of the semigroup $S$.

\begin{lem}\label{Lemma3.4}
The logarithm height function $h$ on $\mathbb{P}^{d}(\bar{\mathbb{Q}})$
satisfies 
\begin{equation}\label{heightS}
h(\phi x)=nh(x), \  \  \forall\phi=n\phi_{0}\in S, \  \phi_{0}\in G.
\end{equation}
\end{lem}

\proof
This is clear since $\phi=n\phi_{0}$ acts on $\mathbb{P}^{d}(\bar{\mathbb{Q}})$
by permuting the homogeneous coordinates according to $\phi_{0}$
and then raising them to $n$-th power.
\endproof

We then obtain the following construction of a time evolution on the endomotives.

\begin{lem}\label{HendoPd} 
Let $h$ be the logarithm height functions on $\P^d$.
For all $t\in {\mathbb R}$, setting
\begin{equation}\label{sigmatPd1}
\sigma_{t}(\mu_{\phi})=n^{it}\mu_{\phi}, \ \  \forall\, \phi=n\phi_{0}\in S \, \text{ with } \, 
\phi_{0}\in G, \ \ \  \text{ and } \ \ \ \sigma_{t}(e(f_{x}))=e(f_{x})
\end{equation}
for all 
$$ f_{x} \in\underset{x\in X_{0}}{\bigsqcup}
{\rm Hom}_{\mathbb{Z}}(\mathbb{Z}{}^{\ell(x)+1}/(1,...,1),\mathbb{Q}/\mathbb{Z}), $$
determines time evolutions on the algebras $\cA_{X,X_0}$ and $\cA_{Y,X_0}$ of \eqref{AX0sum}
and \eqref{AX0prod}. 
The Hamiltonians implementing these time evolutions in the representations of
$\cA_{X,X_0}$ and $\cA_{Y,X_0}$ on the respective Hilbert spaces
$\ell^2(\mathcal{B}_{X,\alpha})$ and $\ell^2(\mathcal{B}_{Y,\alpha})$
are determined in both cases by the assignment, for $\alpha^{k}x^{k'}\in\mathcal{B}_{\alpha,x}$,
\begin{equation}\label{Halphax}
H\epsilon_{\alpha^{k}x^{k'}}=\log(h(\alpha^{k}x^{k'}))\,\, \epsilon_{\alpha^{k}x^{k'}}.
\end{equation}
\end{lem}

\medskip

\proof By Lemma \ref{Lemma3.4} we see that the logarithmic height function satisfies the
condition required for Proposition \ref{Lemma2.6}, hence the operator $H$ defines as in
\eqref{Halphax} is indeed the Hamiltonian of the time evolution, when extended to
an operator on the direct sum, respectively the tensor product, of the spaces
$\ell^2(\mathcal{B}_{\alpha,x})$, for $x\in X_0$.
\endproof

\medskip
\subsubsection{Partition function and a logarithmic height zeta function}

We obtain the height zeta function of (multi)-sets $\mathcal{B}_{X,\alpha}$
and $\mathcal{B}_{Y,\alpha}$ of algebraic points in $\mathbb{P}^{d}(\bar{\mathbb{Q}})$ 
as the partition function of the quantum statistical mechanical system.

\begin{lem}\label{ZetaHeight}
The partition function of the quantum statistical mechanical
system $(\cA_{X,X_0},\sigma_t)$, represented on the
Hilbert space $\ell^2(\mathcal{B}_{X,\alpha})$ is a zeta function of the form
\begin{equation}\label{heightzeta}
Z_{X,X_0,\alpha}(\beta)=\underset{x\in\mathcal{B}_{X,\alpha}}{\sum} h(x)^{-\beta}.
\end{equation}
The multiplicative case is analogous.
\end{lem}

\proof This follows directly from Lemma \ref{HendoPd}, 
and the Hamiltonian \eqref{Halphax} 
that the partition function of the system is given by 
\begin{equation}
Z_{X,X_0,\alpha}(\beta)={\rm Tr}(e^{-sH})=\underset{x\in\mathcal{B}_{X,\alpha}}{\sum}h(x)^{-\beta},
\end{equation}
which is a logarithmic height zeta function of the (multi)-set
$\mathcal{B}_{X,\alpha}$ of algebraic points in $\mathbb{P}^{d}(\bar{\mathbb{Q}})$.
\endproof

\smallskip
\subsubsection{Convergence}

One knows that the exponential height zeta function converges for all $\Re(\beta)>a$,
for some sufficiently large $a>0$. For a zeta function based on the logarithmic height
to have similar convergence properties, one needs the size of the points with
a given height bound inside the sampling set $\mathcal{B}_{X,\alpha}$ to be growing
with only at most logarithmic speed rather than polynomially. 

\smallskip

Lemma \ref{Lemma3.4}, with the scaling property \eqref{heightS} of the height
function, together with the construction of the sets $\cB_{X,\alpha}$ and $\cB_{Y,\alpha}$,
and the action of the semigroup on it as in \eqref{phiBalpha}, show that this is indeed
the case.
 
\smallskip
\subsubsection{Height and degree bounds}

Lemma 3.3 implies that for any $a,b\in\mathbb{R}^{+}$ we
can use $$X_{0}=\{x\in\mathbb{P}^{d}(\bar{\mathbb{Q}})|h(x)\leq a,[\mathbb{Q}(x):\mathbb{Q}]\leq b\}$$
as the starting point of the construction since $G$ acts on it. In
this case the endomotive system carries some of the arithmetic information
on the $\mathbb{\bar{Q}}$-algebraic points in the projective space
with bounded height and degree.

\medskip
\subsection{Endomotives of affine spaces}\label{AdSec}

We modify the construction given above for projective spaces, to
obtain endomotives associated to the affine spaces $\mathbb{A}^{d}$, 
with similar properties.

\smallskip

Note that as a toric variety $\mathbb{A}^{d}$ is given by the fan
spanned by the canonical basis vectors $e_{1},...,e_{d}$ in $N=\mathbb{Z}^{d}$,
and it is easy to see that the group $G$ is given by permutations of cartesian
coordinates in the affine space.

\smallskip

As in Section \ref{PdSec}, we start with a $G$-invariant finite subset $X_{0}$
of $\mathbb{A}^{d}(\bar{\mathbb{Q}})$. Here we need the additional
assumption that for each $x\in X_{0}$, each coordinate of $x$ is
a non-torsion element of $\mathbb{\bar{Q}}^{*}$. Then, we use the same
preimage projective limit construction. The main difference
is that now we are not working with homogeneous coordinates anymore,
so that each point has a unique $d$-tuple coordinate corresponding
to it, which simplifies the construction. The analog of Lemma \ref{Lemma3.1}
now takes the following form.

\smallskip

\begin{lem}\label{Lemma3.5}
In the case of the affine space $\A^d$, 
the algebra $C(\varprojlim_n (nI)^{-1}(x))$
is isomorphic to $C^{*}({\rm Hom}_{\mathbb{Z}}(\mathbb{Z}{}^{\ell},\mathbb{Q}/\mathbb{Z}))$,
where $\ell=\ell(x)$ is the number of nonzero coordinates of $x=(x_{1},...,x_{d})$.
\end{lem}

\proof 
The only difference, with respect to the analogous statement for $\P^d$, 
is that now we do not need to quotient the lattices by $(1,\ldots,1)$.
This, the commutative diagram \eqref{diaglims} now becomes 
\begin{equation}\label{diag2}
\xymatrix{}
\xymatrix{(n_{1}n_{2}I)^{-1}(x)\ar[d]\ar[r]^{\ n_{2}I} & (n_{1}I)^{-1}(x)\ar[d]\\
(\mathbb{Z}/n_{1}n_{2}\mathbb{Z})^{\ell}\ar[r]^{\ n_{2}} & 
(\mathbb{Z}/n_{1}\mathbb{Z})^{\ell}\ensuremath{}.
}
\end{equation}
The rest of the proof follows exactly as in Lemma \ref{Lemma3.1}.
\endproof

\medskip
\subsubsection{The endomotive algebra}

The additive endomotive associated to $\A^d$ with the choice of a finite set $X_0$
of $\mathbb{\bar{Q}}$-algebraic points is the crossed product algebra
\begin{equation}\label{AX0sumAd}
(\oplus_{x\in X_0} C^{*}({\rm Hom}_{\mathbb{Z}}(\Z^{\ell(x)},
\mathbb{Q}/\mathbb{Z})))\rtimes_\rho S
\end{equation}
and the multiplicative endomotive
\begin{equation}\label{AX0prodAd}
(\otimes_{x\in X_0} C^{*}({\rm Hom}_{\mathbb{Z}}(\Z^{\ell(x)},
\mathbb{Q}/\mathbb{Z})))\rtimes_\rho S.
\end{equation}

\medskip
\subsubsection{The representation}

The main difference, with respect to the analogous construction
for projective spaces, appears in the Hilbert space we represent the
algebra on.

\begin{lem}\label{Lemma3.6}
Consider the set 
\begin{equation}
\mathcal{B}_{x}=\{(x_{1}^{k_{1}},...,x_{\ell}^{k_{\ell}})\in\mathbb{A}^{d}(\mathbb{\bar{Q}})\, | \,
k_{1},...,k_{\ell}\in\mathbb{Z}\}.
\end{equation}
Then there is a natural representation of the $C^{*}$-algebra 
$C^{*}({\rm Hom}_{\mathbb{Z}}(\mathbb{Z}{}^{\ell},\mathbb{Q}/\mathbb{Z}))$,
with $\ell=\ell(x)$, for a point $x\in X_0$, on the Hilbert space $\ell^{2}(\mathcal{B}_{x})$,
given by 
\begin{equation}
e(f)\epsilon_{x^{k}}=\exp(2\pi if(k))\, \epsilon_{x^{k}},\ \ \ 
\forall f\in {\rm Hom}_{\mathbb{Z}}(\mathbb{Z}{}^{\ell(x)},\mathbb{Q}/\mathbb{Z}), \text{ and }
\forall k\in\mathbb{Z}^{\ell(x)},
\end{equation}
where we abbreviate $\mbox{(\ensuremath{x_{1}^{k_{1}}},...,\ensuremath{x_{\ell}^{k_{\ell}}})}$
with the notation $x^{k}$.
\end{lem}

The proof is again straghtforward and thus ommited. Note that since
each $x_{j}$ is non-torsion, $\mathcal{B}_{x}$ naturally identifies
with $\mathbb{Z}{}^{\ell}$. Also note that, in this affine case, 
we do not need to introduce a new parameter $\alpha$ to 
define the representation.

\smallskip

The representation of $S$ on the Hilbert spaces $\ell^2 (\mathcal{B}_X)$ and
$\ell^2(\cB_Y)$ where 
$$\mathcal{B}_X=\underset{x\in X_{0}}{\bigsqcup}\mathcal{B}{}_{x} \ \ \ \text{ and }
\ \ \  \mathcal{B}_Y=\underset{x\in X_{0}}{\prod}\mathcal{B}{}_{x} $$
is the same as before, given by the action of $S$ on $\mathcal{B}_X$ and $\cB_Y$,
and with the adjoints given by partial inverses. 

\smallskip

Lemma \ref{Lemma3.3} relating these
operators also holds. This allows us to define the endomotive as the
semigroup crossed product in the usual way, with the same time evolution
and Hamitonian as in Section \ref{PdSec} (note that Lemma \ref{Lemma3.4} also holds in
the affine case).

\medskip
\subsubsection{A logarithmic height zeta function}

The partition function of this system is then given
by a logarithmic height zeta function on $\mathcal{B}_X$ or $\cB_Y$ of the form
\begin{equation}\label{heightZAd}
Z(\beta)=\underset{x\in\mathcal{B}_X}{\sum}h(x)^{-\beta}.
\end{equation}

\section{Gibbs states}\label{GibbsSec}

We consider here again the general construction of endomotives
of abstract toric varieties described in Section \ref{EndoSec}. Thus,
we consider $C^*$-dynamical systems $(\cA_{X,\Sigma},\sigma_t)$
and $(\cA_{Y,\Sigma},\sigma_t)$ as above, with the time evolution
and covariant representations constructed as in Proposition \ref{Lemma2.6}.

\smallskip

\begin{lem}\label{convergebetac}
Let $g: S \to \R^*_+$ be a semigroup homomorphism as in Proposition \ref{Lemma2.6}, 
with functions $h_k$ as in \eqref{hkghk}. Suppose that there is a $\beta_g>0$ such
that for all $\Re(\beta)>\beta_g$ the zeta function
\begin{equation}\label{Zgbeta}
Z_g(\beta)=\sum_{\phi\in S} g(\phi)^{-\beta} < \infty.
\end{equation}
Then there is a choice of $h_k$ as in \eqref{hkghk} such that the zeta function
\begin{equation}\label{Zkbeta}
 Z_k (\beta)= \sum_{f_k \in {\rm Hom}_\Z(\sigma_k^\perp \cap M, \Z)} h_k(f_k)^{-\beta} <\infty 
\end{equation} 
for all $\Re(\beta)>\beta_c$.
\end{lem}

\proof
Let $\cF_{k,S} \subset {\rm Hom}_\Z(\sigma_k^\perp \cap M, \Z)$ be a fundamental
domain for the action of the semigroup $S$. A function $h_k$ 
as in Proposition \ref{Lemma2.6} is determined by assigning $h_k$ on $\cF_{k,S}$
and by extending it using  \eqref{hkghk}.  In particular, one can choose the
function $h_k$ so that, for $\Re(\beta) >\beta_g$ the series
\begin{equation}\label{Zhkbeta}
Z_{\cF_{k,S}}(\beta)=\sum_{f\in \cF_{k,S}} h_k(f)^{-\beta} <\infty.
\end{equation}
By \eqref{hkghk}, the partition function \eqref{Zkbeta} factors as a product
$$ Z_k (\beta)= Z_g(\beta) \,\, Z_{\cF_{k,S}}(\beta), $$
hence the statement follows.
\endproof

\smallskip

The previous statement can be reformulated as in Corollary \ref{Zcor},
in the case where $S=\N \times G$ with $G\subset {\rm Gl}_d(\Z)$.

\medskip
\subsection{Polylogarithm-type functions on toric varieties}

We now consider Gibbs states of the form
\begin{equation}\label{Gibbs}
\varphi(a) = \frac{{\rm Tr}(\pi(a) \, e^{-\beta H})}{{\rm Tr}(e^{-\beta H})},
\end{equation}
in a given covariant representation $(\pi, H)$ of a $C^*$-dynamical
system $(\cA,\sigma_t)$. These are well defined and satisfy the KMS$_\beta$ 
condition whenever ${\rm  Tr}(e^{-\beta H}) < \infty$.

\smallskip

In the case of the original Bost--Connes system, the Gibbs states \eqref{Gibbs}
are well defined for $\beta >1$ and they are values at roots of unity $\zeta_r$, $r\in \Q/\Z$, 
$$ \varphi(e(r)) =\zeta(\beta)^{-1} \sum_{n\geq 1} \frac{\zeta_r^n}{n^\beta}
=\zeta(\beta)^{-1} {\rm Li}_\beta(\zeta_r) $$
of the polylogarithm function
$$ {\rm Li}_s(z) =\sum_{n=1}^\infty \frac{z^n}{n^s}, $$
normalized by the Riemann zeta function. These states depend on the
choice of an element $\rho\in \hat\Z^*$ viewed as a choice of an embedding 
of the abstract roots of unity $\Q/\Z$ in $\C$, with $\zeta_r=\rho(r)$.

\smallskip

Gibbs states of endomotives of toric varieties
provide an analog of polylogarithm functions on toric varieties, of the form
\begin{equation}\label{polyGibbs}
\varphi(e(\underline{r}))=\frac{1}{Z(s)} \,\, \sum_{k=1}^m 
\sum_{f_k \in {\rm Hom}_\Z(\sigma_k^\perp\cap M, \Z)} \exp(2\pi i f_k(r_k)) \,\, h_k(f_k)^{-\beta},
\end{equation}
normalized by the zeta function
$$ Z(s) = \sum_{k=1}^m \sum_{f_k \in {\rm Hom}_\Z(\sigma_k^\perp\cap M, \Z)}  h_k(f)^{-\beta} . $$
These are KMS$_\beta$ states for the system $(\cA_{X,\Sigma}, \sigma_t)$. The case of
$\cA_{Y,\Sigma}$ is similar. As in the case of the Bost--Connes system, we can change the
KMS-state \eqref{polyGibbs} by precomposing with an automorphism of the dynamical
system in ${\rm Hom}_\Z(\sigma_k^\perp\cap M, \hat\Z^*)$, as in Lemma \ref{auto}.

\bigskip

\subsection*{Acknowledgments}
This paper is based on the results of the first author's summer research project,
supported by a Summer Undergraduate Research Fellowship at Caltech. The
second author acknowledges support from NSF grants
DMS-0901221, DMS-1007207, DMS-1201512, PHY-1205440.

\end{document}